\documentclass[runningheads]{llncs}

\usepackage[T1]{fontenc}
\usepackage[hyphens]{url}
\usepackage{graphicx}
\usepackage[numbers]{natbib}

\usepackage{amsmath}
\usepackage{algorithm}
\usepackage{algpseudocode}
\usepackage{booktabs}
\usepackage[hidelinks]{hyperref}

\title{The Montparnasse Algorithm for RNA Design}

\author{Tristan Cazenave}

\institute{LAMSADE, Universit\'e Paris Dauphine - PSL, CNRS, Paris, France\\
\email{Tristan.Cazenave@dauphine.psl.eu}}

\begin{document}

\maketitle

\begin{abstract}
Inverse RNA Folding consists of finding a nucleotide sequence that folds into
a predefined secondary structure. It is useful for synthetic biology,
medicine, and nanotechnology. We propose Montparnasse, a Monte Carlo search
framework based on Generalized Nested Rollout Policy Adaptation, augmented
with a structural prior derived from the energy model, slow and long
adaptation at level~1, a lexicographic multicriteria evaluation, a local
repair of the pairs the predicted folding gets wrong, and a fleet of parallel
workers coupled through a shared beam whose diversity is maintained on the
\emph{faults} the candidates make rather than on the candidates themselves.
On the Eterna100~V1 benchmark, with both solvers folding through the same
ViennaRNA~1.8.5 library under the Turner~1999 model, with 100 parallel workers
and a budget of 30 minutes per puzzle, Montparnasse solves a median of 96 of
the 100 puzzles against a median of 90 for DesiRNA, the previous state of the
art, each figure taken over five passes. It is faster on 88 of the 90 puzzles
that both solve, with a median solving time of 0.9~s against 5.6~s, and it
solves 55 puzzles within the first second. The one puzzle no pass reaches is solved under the same
model with a longer budget, so every Eterna100~V1 target is solved by
Montparnasse.

\keywords{Monte Carlo Search \and Nested Rollout Policy Adaptation \and
Inverse RNA Folding \and RNA Design \and Parallel Search}
\end{abstract}

\section{Introduction}

The design of molecules with specific properties is an important research
topic for health-related applications. The Inverse RNA Folding problem, which
consists of finding an RNA molecule that folds into a predefined shape, is a
difficult combinatorial problem. It is important for scientific fields such as
bioengineering, pharmaceutical research, biochemistry, synthetic biology, and
RNA nanostructures \citep{portela2018unexpectedly}. RNA is involved in many
biological functions, synthetic RNA can be easily produced \citep{Reese2005},
and RNA design has many applications in synthetic biology and drug design,
including the construction of riboswitches and ribozymes.

RNA molecules are long molecules composed of four possible nucleotides and can
be represented as strings over the alphabet $\{A, C, G, U\}$ (Adenine,
Cytosine, Guanine, Uracil). For a molecule of length $N$, the size of the
sequence space is $4^N$, which is intractable for long molecules. The sequence
folds back on itself to form its secondary structure, which can be predicted
from the sequence in polynomial time. The opposite direction, the Inverse RNA
Folding problem, is hard \citep{bonnet2020designing}. RNA function is
determined by the tertiary structure, which is itself largely determined by
the secondary structure through base-pairing interactions. The bonds between
two nucleotides are given by the six possible base pairs (CG, GC, AU, UA, UG,
GU). The dot-bracket notation is used to represent the secondary structure:
matching opening and closing brackets denote base pairs, and dots denote
unpaired positions.

This paper presents Montparnasse, a Monte Carlo search framework for Inverse
RNA Folding based on Generalized Nested Rollout Policy Adaptation (GNRPA).
Montparnasse augments GNRPA with five components: a structural prior on the
policy, a slow and long adaptation regime at level~1, a lexicographic
multicriteria evaluation of candidate sequences, a local repair of the pairs
the predicted folding gets wrong, and a fleet of parallel workers coupled
through a shared beam. Eterna \citep{eterna} is a standard benchmark for
Inverse RNA Folding algorithms; many algorithms have been applied to it over
the years, with DesiRNA \citep{wirecki2025desirna} recently achieving full
coverage. Under a budget of 30 minutes per puzzle and with both solvers
folding through the same library, Montparnasse solves a median of 96 of the
100 Eterna100 V1 puzzles against a median of 90 for DesiRNA, five passes on
each side, and the one no pass reaches is solved under the same model with a
longer budget.

The part of the algorithm that is new in this paper is the fleet. The usual
way to parallelize a Monte Carlo search is to run independent searches and to
stop as soon as one succeeds. On the hard puzzles this stops paying very
early, because the workers rediscover one another's partial designs. The fleet
of Montparnasse couples its workers through a shared beam of the best
sequences found so far, and the rule that admits a sequence into that beam is
stated on the \emph{faults} of its predicted folding rather than on its
nucleotides. The measurements that motivate each rule, and the controls that
refute the rules we discarded, are reported along with it.

The paper is organized as follows. Section~\ref{sec:previous} reviews previous
work on RNA design. Section~\ref{sec:montparnasse} presents the Montparnasse
algorithm. Section~\ref{sec:fleet} presents the fleet.
Section~\ref{sec:experiments} reports experimental results on Eterna100~V1.
The last section concludes and outlines future work.

\begin{figure}[t]
  \centering
  \includegraphics[width=\linewidth]{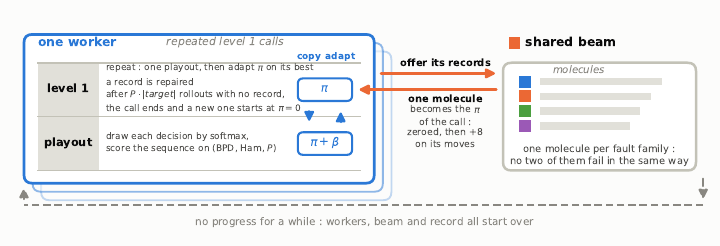}
  \caption{Montparnasse in one picture. A fleet of a hundred workers searches
  the same target at once, each repeating level~1 calls. A call holds a policy
  $\pi$, passes a copy of it to each playout, which draws every decision by
  a softmax over $\pi$ and the structural prior $\beta$, and adapts it on
  the best sequence returned, until $P \cdot |target|$ rollouts have passed
  with no record. The workers are coupled by a shared beam holding one
  molecule per fault family, so that no two of them fail in the same way. A
  worker offers its records to the beam and takes one molecule back, and what
  that molecule becomes is the policy of its call: $\pi$ is zeroed and every
  move of the molecule receives a boost.}
  \label{fig:schema}
\end{figure}

\section{Previous Work}
\label{sec:previous}

This section reviews work relevant to Montparnasse: earlier approaches to
Inverse RNA Folding, GREED-RNA, DesiRNA, and Monte Carlo search.

\subsection{Earlier Approaches to Inverse RNA Folding}

Inverse RNA Folding has been addressed by adaptive random walk in
RNAinverse~\citep{hofacker1994fast}, stochastic local search in
RNA-SSD~\citep{andronescu2004new} and INFO-RNA~\citep{busch2006info}, and
evolutionary algorithms such as MODENA~\citep{taneda2015multi} and
aRNAque~\citep{merleau2022arnaque}. The recent RNA design
book~\citep{churkin2024rna} surveys methods for both secondary and tertiary
structure design.

\subsection{GREED-RNA}

GREED-RNA~\citep{lozano2024simple} is a recent local search method for Eterna.
It initializes all base pairs as GC or CG and all unpaired positions as A. In
the early phase of the search it applies greedy mutations that swap GC and CG
at random positions; later it switches to random mutations over all base pairs
and nucleotides. Candidates are ranked by a multi-objective evaluation that
uses, in order, base-pair distance, Hamming distance, ensemble probability,
partition function, ensemble defect, and GC-content distance. Restarts are
triggered when a stagnation counter reaches a threshold, drawing from a pool
of sorted sequences. Montparnasse borrows the idea of a lexicographic
multicriteria evaluation from GREED-RNA.

\subsection{DesiRNA}

DesiRNA~\citep{wirecki2025desirna} targets specified secondary structures,
including pseudoknots, RNA--RNA complexes, alternative conformational states,
and prevention of unwanted oligomerization. It is based on Replica Exchange
Monte Carlo, a parallel tempering approach in which multiple simulations run
at different temperatures and periodically exchange configurations to balance
global exploration with local refinement. The fitness function minimizes the
difference between the free energy of the full thermodynamic ensemble and that
of the target structure, accounting for the entire Boltzmann distribution
rather than only the MFE structure. DesiRNA has been benchmarked on
Eterna100~V1 and used to design functional \textit{glmS} ribozyme variants
whose catalytic activity was experimentally validated. It is the baseline we
compare against.

\subsection{Monte Carlo Search}

Early work combined UCT with local search for RNA design~\citep{yang2017rna}
but was not evaluated on Eterna. NEMO uses Nested Monte Carlo
Search~\citep{CazenaveIJCAI09} with a handcrafted rollout policy and local
search refinement, and solves 95 of the 100 Eterna V1 puzzles. Generalized
Nested Rollout Policy Adaptation (GNRPA)~\citep{Cazenave2020GNRPA} was applied
to Inverse RNA Folding~\citep{Cazenave2020Inverse}, also reaching 95 puzzles,
and was later refined by learning a prior for the policy from
transformers~\citep{cazenave2024monte} or from statistics on solved
problems~\citep{cazenave2024learning}. GNRPA with Limited Repetitions
(GNRPALR)~\citep{cazenave2024generalized} avoids overly deterministic policies
by stopping iterations when the same sequence is sampled twice at a given
level.

\begin{algorithm}[t]
\caption{Rollout for Inverse RNA Folding}
\label{alg:rollout}
\begin{algorithmic}[1]
\Function{Rollout}{$\pi$, $\mathit{target}$}
\State $x \gets$ array of length $|\mathit{target}|$, all entries unassigned
\State $\mathit{pairs} \gets \textsc{MatchBrackets}(\mathit{target})$
\For{$i = 0$ \textbf{to} $|\mathit{target}| - 1$}
    \If{$x_i$ already assigned} \Comment{closing a pair}
        \State \textbf{continue}
    \EndIf
    \If{$\mathit{target}_i = \textrm{``.''}$}
        \State $A_i \gets \{A, C, G, U\}$
    \Else \Comment{$\mathit{target}_i = \textrm{``(''}$}
        \State $A_i \gets \{GC, CG, AU, UA, GU, UG\}$
    \EndIf
    \For{each $a \in A_i$}
        \State $z_a \gets \pi_i(a) + \beta_i(a)$
    \EndFor
    \For{each $a \in A_i$}
        \State $p_a \gets e^{z_a} / \textstyle\sum_{a' \in A_i} e^{z_{a'}}$
    \EndFor
    \State $a^* \gets$ sample from $A_i$ according to $(p_a)_{a \in A_i}$
    \If{$\mathit{target}_i = \textrm{``.''}$}
        \State $x_i \gets a^*$
    \Else
        \State $j \gets \mathit{pairs}[i]$
        \State $(n_{\text{open}}, n_{\text{close}}) \gets a^*$
        \State $x_i \gets n_{\text{open}}$; $x_j \gets n_{\text{close}}$
    \EndIf
\EndFor
\State \Return $x$
\EndFunction
\end{algorithmic}
\end{algorithm}

\begin{algorithm}[t]
\caption{Adapt}
\label{alg:adapt}
\begin{algorithmic}[1]
\Function{Adapt}{$\pi$, $x^*$, $\alpha$}
\State $\pi' \gets \pi$ \Comment{copy of the policy}
\For{$i = 0$ \textbf{to} $N - 1$}
    \If{position $i$ has no action in $x^*$}
        \State \textbf{continue}
    \EndIf
    \State $A_i \gets$ set of legal actions at position $i$
    \State $a^* \gets$ action taken by $x^*$ at position $i$
    \State $\pi'_i(a^*) \gets \pi'_i(a^*) + \alpha$
    \For{each $a \in A_i$}
        \State $z_a \gets \pi_i(a) + \beta_i(a)$
    \EndFor
    \State $Z \gets \sum_{a' \in A_i} e^{z_{a'}}$
    \For{each $a \in A_i$}
        \State $p_a \gets e^{z_a} / Z$
        \State $\pi'_i(a) \gets \pi'_i(a) - \alpha \cdot p_a$
    \EndFor
\EndFor
\State \Return $\pi'$
\EndFunction
\end{algorithmic}
\end{algorithm}

\section{Montparnasse}
\label{sec:montparnasse}

This section presents the Montparnasse algorithm, which is a combination of
NRPA \citep{Rosin2011} with a prior \citep{Cazenave2020GNRPA}, a lexicographic
multicriteria evaluation \citep{lozano2024simple}, a slow and long adaptation
at level~1, a local repair of the faulty pairs, and the fleet of
Section~\ref{sec:fleet}.

In Montparnasse, each move is associated with a weight stored in an array
called the policy. For each level, there is a best sequence and a policy. The
principle is to reinforce the weights of the best sequence of moves found
during the iterations at each level. At the lowest level, the weights are used
in the softmax function to produce a rollout policy that generates good
sequences of moves.

NRPA \citep{Rosin2011} uses nested search \citep{CazenaveIJCAI09}. At each
level, it takes a policy as input and returns a sequence and its associated
scores. At any level $>$ 0, the algorithm makes numerous recursive calls to
the lower level, adapting the policy each time with the best sequence of moves
to date. The changes made to the policy do not affect the policy in higher
levels. At level 0, NRPA returns the sequence obtained by the rollout function
as well as its associated scores.

The rollout function sequentially constructs a random solution biased by the
policy weights until it reaches a terminal state. At each position, it
performs Boltzmann sampling, choosing actions with probabilities given by the
softmax function.

Let $\pi_i(a)$ be the weight associated with action $a$ at position $i$ in the
policy. In NRPA, the probability of choosing action $a$ at position $i$ is:
$$ p_i(a) = \frac{e^{\pi_i(a)}}{\sum_{a' \in A_i} e^{\pi_i(a')}} $$
where $A_i$ is the set of legal actions at position $i$.

GNRPA \citep{Cazenave2020GNRPA} generalizes the softmax with a per-action bias
$\beta_i(a)$. The probability of choosing action $a$ at position $i$ becomes:
$$ p_i(a) = \frac{e^{\pi_i(a) + \beta_i(a)}}{\sum_{a' \in A_i} e^{\pi_i(a') +
\beta_i(a')}} $$

Setting $\beta_i(a) = 0$ for all $i$ and $a$ recovers the NRPA formula, which
corresponds to sampling without prior.

Montparnasse uses a single level of that nesting, and
Algorithm~\ref{alg:montparnasse} is what one worker runs. A level~1 call
starts from a null policy, repeats a rollout followed by an adaptation on the
best sequence of the call, and gives up when the patience counter reaches
$P \cdot |\mathit{target}|$ rollouts without an improvement, with the patience
factor $P = 4$. The worker then starts another call. The learning rate is
0.1 rather than the 1.0 of NRPA, which slows specialization and keeps the
diversity of the solutions seen at this level.

There is no level above, and the measurements are what removed it. The
recursion we first ran made one level~1 call per iteration of a level~2 that
adapted a policy of its own; over a full Eterna100~V1 run the global record
was set during the \emph{first} level~1 call on 97 of the 99 puzzles solved,
the exceptions being puzzle~85 on the second call and puzzle~87 on the fourth,
while the worker that went furthest reached eight calls in 78 minutes. What
carries a good position from one call to the next is not a policy learned
above them but the sequence itself, through the beam of
Section~\ref{sec:fleet}: every candidate of a call is offered to the beam, and
every call starts on a molecule taken back from it.

Improvement is assessed lexicographically: a new solution $x$ with score $s$
replaces the current best whenever $s < s^*$ under the lexicographic ordering,
which prioritizes the primary objective and breaks ties by the following ones.
If no improvement is found, a patience counter is incremented;
otherwise, it is reset to zero. The policy adaptation step
$\textsc{Adapt}(\pi, x^*, \alpha)$ increases the weight of each action taken
by $x^*$ by $\alpha$ and subtracts $\alpha$ times the softmax probability from
all possible actions, reinforcing the best solution found so far by the
call.

A transposition table is used to cache the secondary structure of RNA
molecules that have already been folded previously during the current search.

\begin{algorithm}[t]
\caption{Montparnasse, one worker}
\label{alg:montparnasse}
\begin{algorithmic}[1]
\Function{Montparnasse}{$\mathit{target}$}
\Loop \Comment{one level 1 call per iteration}
\State $\pi \gets 0$ \Comment{a call starts from a null policy}
\State $x^* \gets \text{None}$;\ \ $s^* \gets (+\infty, \dots, +\infty)$
\State $\mathit{patience} \gets 0$;\ \ $k \gets \text{None}$
\While{$\mathit{patience} < P \cdot |\mathit{target}|$}
\State $k, \pi \gets \textsc{Follow}(k, \pi)$ \Comment{the policy only}
\State $x \gets \textsc{Rollout}(\pi)$;\ \ $s \gets \textsc{Score}(x)$
\State \textsc{Offer}($x$, $s$) \Comment{Algorithm~\ref{alg:beam}}
\If{$s < s^*$} \Comment{lexicographic}
\State $x^* \gets \textsc{Repair}(x)$;\ \ $s^* \gets \textsc{Score}(x^*)$
\State $\mathit{patience} \gets 0$
\If{$s^*_1 = 0$} \State \Return $x^*$ \Comment{the target is folded}
\EndIf
\Else
\State $\mathit{patience} \gets \mathit{patience} + 1$
\EndIf
\State $\pi \gets \textsc{Adapt}(\pi, x^*, 0.1)$
\EndWhile
\EndLoop
\EndFunction
\end{algorithmic}
\end{algorithm}

\subsection{Rollout}

The rollout function for Inverse RNA Folding
(Algorithm~\ref{alg:rollout}) constructs a candidate sequence by traversing
the target dot-bracket structure from left to right. The action space at each
position depends on the target symbol: an unpaired position (``.'') admits the
four single nucleotides $\{A, C, G, U\}$, while an opening bracket (``('')
admits the six base pairs $\{GC, CG, AU, UA, GU, UG\}$. Selecting a base pair
at an opening bracket simultaneously assigns the nucleotide at the matching
closing bracket, whose index is precomputed by $\textsc{MatchBrackets}$;
closing positions are therefore skipped during the traversal. At each
undetermined position, the action is sampled by softmax over the sum $\pi_i(a)
+ \beta_i(a)$, where $\pi_i$ is the learned policy weight at position $i$ and
$\beta_i$ is the prior.

\subsection{Multicriteria Evaluation}
\label{sec:multi}

The candidate is folded and the predicted structure is compared to the target
with a lexicographic multicriteria evaluation, in this order: base-pair
distance, Hamming distance between the predicted and target structures,
probability of the target structure in the Boltzmann ensemble, ensemble defect
with respect to the target, and partition function free energy. The first
criterion is the one the benchmark is scored on; the others only break its
ties, and they are what makes a plateau at a constant base-pair distance
informative instead of flat.

\subsection{Prior}
\label{sec:prior}

The uniform initial policy of NRPA carries no structural information.
Montparnasse uses a prior derived from the energy model and from the topology
of the target.

At paired positions, $\beta(GC) = \beta(CG) = 4$ and $\beta(AU) = \beta(UA) =
\beta(GU) = \beta(UG) = 0$, which puts GC and CG at 48\% each and favours the
more thermodynamically stable pairs without making the choice deterministic. A
much larger value such as $\beta = 11$, used in earlier work, leaves the four
other pairs a combined probability below $10^{-4}$ and puts the AU-rich
and wobble-rich designs that some targets need out of reach of the rollout.

At unpaired positions, the base prior is $\beta(A) = 5$, $\beta(C) = 1$,
$\beta(G) = 3$, $\beta(U) = 0$, biasing toward adenine, which is less prone to
forming spurious base pairs in loop regions. On top of it, a positional prior
of strength $S = 2$ is computed once from the topology of the target: A and G
at the edges of an interior loop, G and U inside a $2 \times 2$ interior loop,
and the GNRA motif in a tetraloop.

The last term of the prior is a bonus of $1.20\,S$ on G at the \emph{first}
base of a hairpin loop of five bases or more, and nothing at the last base. It
was obtained by exhaustively enumerating the $4^5$ sequences of a five-base
hairpin loop under the energy model used in the experiments. Conditioned on
the far end of the loop being A, which is what the unpaired prior above
already imposes, the Boltzmann marginal of G at the first base is 59.3,
52.4, 57.4 and 64.8\% under closing pairs GC, CG, AU and UA respectively, and
G is worth $-1.30$ kcal/mol under a GC closure. The bonus places G at exactly
that marginal. The symmetric version of this prior, which we used first, is
worse than nothing: conditioned on the first base being A, the marginal of G
at the \emph{last} base of the loop falls to between 14.8 and 30.2\%, below
its own unconditioned marginal of between 28.0 and 46.9\%, so a bonus there
pushes the rollout away from the distribution of the model instead of towards
it. The prior is computed once at initialization and remains fixed across
iterations; only $\pi$ is updated by \textsc{Adapt}.

\subsection{Slow and Long Adaptation at Level 1}

Standard NRPA at level 1 uses a fixed learning rate $\alpha = 1.0$ and 100
iterations. Montparnasse uses $\alpha = 0.1$ and the patience rule above: the
reduced learning rate slows policy specialization, while late stopping buys
more search for each improvement. On a target of 100 nucleotides the level
runs 400 rollouts without a record before it gives up, against the 100
iterations of plain NRPA.

\subsection{Local Repair}

A sequence that is one pair short of the target is not improved by more
sampling, it is improved by looking at the pair it gets wrong. Whenever a
level~1 call sets a record whose base-pair distance is at most 25,
Montparnasse repairs it. The repair draws at most three of the pairs the
predicted folding gets wrong and, for each of them, builds a context of
\emph{slots}: the faulty pair itself, the target pairs that enclose it and the
target pairs it encloses up to a given depth, and the unpaired positions
adjacent to the ends of all of these. It then climbs on those slots one at a
time and, when the climb sticks, enumerates couples of slots, because some
faults need two simultaneous changes: puzzle 90 needs an AG to GA swap whose
two halves each make the sequence worse, so no one-slot climb reaches it. On
records at base-pair distance 5 or below, the full repair also varies the
faulty pair and two of its loop nucleotides at once.

Both directions of failure are treated the same way. A sequence can fail
because the predicted folding drops a target pair, or because it adds a pair
of its own that the target does not ask for; the second kind has no entry in
the target topology, so its context is built from the positions around it. Of
34 sequences Montparnasse left at base-pair distance 1 in an earlier
measurement, 14 were of the first kind and 20 of the second, which is why the
repair treats both.

\section{The Fleet}
\label{sec:fleet}

Montparnasse runs as a fleet of $W$ worker processes on one machine, with $W =
100$ in the experiments reported here. The naive use of this parallelism is
$W$ independent searches, stopped as soon as one of them succeeds. If $P(t)$
is the probability that one search finds a solution at time $t$, then the
probability that at least one of $W$ independent searches has found it at time
$t$ is
$$P(W,t) = 1 - (1 - P(t))^W.$$
For example, if a search has a probability of 0.2 of solving a problem in one
day, then the same search run on 10 independent threads has a probability $1 -
(1 - 0.2)^{10} \approx 0.89$ of solving it in one day.

This law is a good model of the easy puzzles and a bad model of the hard ones.
On a hard target, the searches do not fail independently: they descend to the
same partial designs and get stuck on the same faults, and the extra workers
buy redundancy instead of coverage. The fleet of Montparnasse therefore
couples its workers through a shared beam, and the admission rule of that beam
is the mechanism by which the workers are kept from converging.

\begin{algorithm}[t]
\caption{The shared beam}
\label{alg:beam}
\begin{algorithmic}[1]
\Function{Offer}{$x$, $s$} \Comment{every candidate of every worker}
\If{the beam has been full and $s \geq$ its worst key} \State \Return
\EndIf
\State $f \gets \textsc{Faults}(\textsc{Fold}(x), \mathit{target})$
\State $e \gets \text{None}$
\If{some rank $j$ has $\textsc{Faults}(j) = f$}
\If{$s \geq \textsc{Key}(j)$} \State \Return \Comment{its family is taken}
\EndIf
\State $e \gets j$ \Comment{$x$ takes its place}
\EndIf
\State $N \gets \{\,j \neq e : x$ differs from rank $j$ on $< D$ helix words$\,\}$
\If{$\exists j \in N : \textsc{Key}(j) \leq s$} \State \Return
\EndIf
\State drop the ranks of $N \cup \{e\}$, insert $(x, s)$, keep the $K$ best
\EndFunction
\Statex
\Function{Follow}{$k$, $\pi$} \Comment{at every iteration of a level 1 call}
\State $j \gets$ this worker's rank, wrapped if the beam is shorter
\If{the beam is empty \textbf{or} $\textsc{Key}(j) = k$} \State \Return $k, \pi$
\EndIf
\State $x_j \gets$ the sequence of rank $j$
\State $\pi \gets 0$ \Comment{a fresh policy, not the call's}
\For{each decision $(i, a)$ of $x_j$ in the rollout order}
\State $\pi_i(a) \gets \pi_i(a) + S$
\EndFor
\State \Return $\textsc{Key}(j), \pi$
\EndFunction
\end{algorithmic}
\end{algorithm}

\subsection{The Beam and Its Followers}

A shared memory segment holds the $K$ best distinct sequences the fleet has
found, one per rank, and Algorithm~\ref{alg:beam} is the two operations a
worker performs on it. Each worker is a \emph{follower}: follower $k$ is given
rank $k$, and \textsc{Follow} replaces the policy of its level~1 call by one
built from that rank's sequence instead of leaving it to the prior alone. When the beam holds fewer ranks than there are
followers, the extra followers wrap around onto the existing ranks. This is
not copying the leader, a single worker holds rank 0, it is spreading
the fleet across the whole front of the search.

The effect is large. On puzzle 99, a fleet of followers reaches base-pair
distance 3 in 28 minutes where a fleet of independent workers needs five
hours, and base-pair distance 1 in about eight hours against 44{,}437~s. The
cost is also known: there is no independent lineage left, and the beam
tightens. On that same target, the median distance between two ranks of the
beam went from 8\% of the length at base-pair distance 4 to 1.1\% at base-pair
distance 1, which is three mutations between fifty sequences. The two
admission rules below are what keeps that from happening.

\subsection{Anchoring a Follower on Its Rank}

A follower is biased towards its seed by adding a constant $S = 8$ to the
logit of every move of the seed sequence, over a freshly zeroed policy, rather
than by iterating \textsc{Adapt} towards it. Iterated adaptation saturates and
anchors nothing on a long target: on 382 nucleotides, three passes of
\textsc{Adapt} leave each move of the seed at a probability between 0.80 and
0.87, so a rollout departs at some 55 positions and never returns. The logit boost is exact: at $S = 4$ a rollout
departs at 63 positions, at $S = 6$ at 13, at $S = 8$ at 3, and at $S = 10$ it
reproduces the seed exactly. We use $S = 8$: the beam holds one rank per
family of faults (next subsection), so a follower searches the sequences that
fail on the same pairs as its seed, and three mutations stay in that set where
thirteen are enough to leave it. The
anchor is only a starting policy: the level~1 call moves away from it through
its own adaptations, and a follower re-reads its rank at every iteration of
that call rather than only when it starts, since a call lasts for hours and
the beam moves under it.

Re-reading replaces the policy and nothing else. The call keeps the best
sequence it has found so far, and the \textsc{Adapt} that closes every
iteration goes on reinforcing that sequence rather than the molecule just
read. A follower that re-hooks is therefore pulled towards two points at once,
the new molecule at a logit of $S$ and its own record at $0.1$ per iteration,
and it adopts the new one only once a rollout drawn from it beats that
record.

The asymmetry is deliberate. A variant that adopts the molecule read as the
record of the call too, at the same budget and worker count, solves 92 of the
hundred against a median of 96. It misses eight: two that all five passes
of Table~\ref{tab:eterna100v1} solve, 73 and 90, five those passes
disagree on among themselves, and puzzle 99, which none of them solves. Time
is the clearer signal: 4{,}784~s on the 92 it still
solves against a median 1{,}959~s, saving 71~s on 51 targets that fall within
seconds and losing 2{,}896~s on 31 others. The record a follower holds
while it sits at base-pair distance~1 is the capital of the search.

\subsection{Diversity on the Faults}

The first admission rule we used required two ranks of the beam to differ by
at least 3\% of the length of the target. It measures the wrong thing.
Measured on puzzle 100 on a plateau at base-pair distance 5, the seven
sequences on the front were 2.9 to 10.5\% apart, the beam reported a median
distance of 7.1\% and held them to be diverse, and they carried exactly the
same list of faults: the four missing pairs $(6,375)$, $(7,374)$, $(15,366)$,
$(16,365)$ and the extra pair $(318,355)$. Twelve workers were working on
twelve variants of one single failure.

Montparnasse replaces the distance rule with a rule on fault families. The
\emph{fault list} of a candidate is the list of target pairs its predicted
folding misses together with the pairs that folding adds, and two candidates
with the same fault list belong to the same family. A candidate whose family
is already held by a rank of the beam is admitted only if it beats that rank,
and it then takes its place. The beam therefore holds one representative per
way of failing.

In a control pair of runs on the same target differing only by this rule, the
run without it reached base-pair distance 5 at 1{,}320~s and was still there
at 20{,}400~s, while the run with it reached 5 at 4{,}200~s, 4 at 4{,}380~s, 3
at 7{,}870~s and 2 at 8{,}890~s. Distance 2 is the best ever reached on that
puzzle, and it was reached in 8{,}890~s against 60{,}615~s for the previous
best. The diversity costs speed early and buys depth later. Putting the
distance rule back on top of the family rule changed nothing: at 1{,}980~s the
run with both rules and the run with the family rule alone were on the same
plateau with the same profile, the beam holding a minimum distance of 2.9\%
against 0.5\% without it. The lever is the diversity of the failures, not the
diversity of the material.

\subsection{Diversity on the Helix Words}

A beam full of distinct fault families can still hold a single design.
Changing one helix of a sequence changes its folding, hence its fault list, so
a hundred one-helix mutants of one design present a hundred distinct families
and satisfy the rule above. This is not a hypothetical: measured on puzzle
100, a beam of 100 ranks held 100 distinct fault families, and yet the median
helix carried the same \emph{word}, the concatenation of the base pairs of
the helix, read along the stem, on 97 of the 100 ranks, and 6 of the 21
helices of the target carried one single word across the entire beam. Those
six helices were frozen: no follower would ever explore them.

Montparnasse therefore adds a second admission rule on top of the family rule:
two ranks of the beam must carry different words on at least $K$ helices, with
$K = 3$ in the experiments of Table~\ref{tab:eterna100v1}. The rule is checked
only where it can be satisfied; on a target with three helices or fewer it is
relaxed, otherwise the beam would collapse to a single rank. Measured at equal
base-pair distance on three puzzles, the number of frozen helices falls from
22 to 9 on puzzle 99, from 20 to 12 on puzzle 90, and from 23 to 1 on puzzle
85.

\subsection{Restarting the Fleet}

A fleet anchored on one beam can descend into a dead end as one. Puzzle 100
does it reproducibly: the fleet reaches base-pair distance 1 between
4{,}500 and 6{,}300~s and then stays there, in one run for more than
125{,}000~s, and the dead end is not the same twice, an extra pair on one
branch in two runs and inside another module in a third. No admission rule
reaches this failure, because the beam is what the whole fleet is anchored
on and the beam is what is stuck.

Montparnasse can therefore restart the whole fleet when its record stops
moving: the workers, their policies, the beam and the record itself are
discarded, and the search begins again from the prior of
Section~\ref{sec:prior}. The rule carries no constant fitted to a target. A
restart fires when the record has been silent for $R$ times the time the
current epoch took to reach it, and for at least $M$ seconds; what descends
fast restarts fast, and a target on which progress continues is never
interrupted. Silence is measured on the whole sorting key of
Section~\ref{sec:multi} and not on the base-pair distance alone, which is
not a detail: on one plateau of puzzle 100 the distance froze at 5 after
622~s while the probability term kept improving by epsilons up to 3{,}642~s,
and a distance-only clock discarded a front of 40 workers after 3{,}600~s of
a silence the key never had.

What the restart buys is insurance, and it is paid for in speed. On puzzle
100, one run of 14.9 hours under this rule divides into eighteen epochs of
median duration 2{,}664~s, which reach base-pair distance 2 or better three
times and distance 1 once, after 8{,}444~s. Five continuous runs of the same
configuration reach distance 1 between 469 and 4{,}484~s three times out of
five, and twice stay at distance 7 for good. A run that is watched and
relaunched by hand is thus better off continuous, and a run left alone is
better off restarting. The floor $M$ only protects young epochs, since past
a certain age the proportional term exceeds it on its own, and an epoch
announces early what it is worth: in that same run the nine epochs that
ended at distance 6 or worse had made their last progress between 205 and
1{,}093~s, and the six that ended at distance 4 or better between 819 and
2{,}768~s.

The restart is off in every run of Table~\ref{tab:eterna100v1} and on in the
run of Table~\ref{tab:dangles2}, where $R = 0$ leaves only the floor, set
there to 1{,}800~s. These measurements are from one target, and we do not
extend them to the benchmark.

\section{Experimental Results}
\label{sec:experiments}

\begin{table}[t]
  \centering\small
  \caption{Eterna100 V1 puzzles solved (out of 100) by Montparnasse with 100
  workers and by DesiRNA with 100 replicas, within each time limit, both
  folding through ViennaRNA 1.8.5 under the Turner 1999 model with
  \texttt{dangles=1}. Each row is the median of five passes of one
  configuration, at the end of the 30-minute budget solving 99, 95, 96, 97 and
  96 for Montparnasse and 91, 89, 91, 89 and 90 for DesiRNA; the difference row
  subtracts them. Montparnasse solves 93 of the hundred in every pass and six,
  76, 85, 94, 95, 97 and 100, in some and not others, all after 140~s;
  puzzle 99 in none.}
  \label{tab:eterna100v1}
  \begin{tabular}{lrrrrrrrr}
    \toprule
    & \multicolumn{8}{c}{Puzzles solved within a time limit (seconds)} \\
    \cmidrule(l){2-9}
    & 10 & 20 & 40 & 80 & 160 & 320 & 640 & 1{,}280 \\
    \midrule
    DesiRNA      & 61 & 67 & 73 & 80 & 87 & 88 & 88 & 89 \\
    Montparnasse & \textbf{79} & \textbf{81} & \textbf{85} & \textbf{87} &
                   \textbf{88} & \textbf{89} & \textbf{93} & \textbf{95} \\
    \midrule
    difference   & $+18$ & $+14$ & $+12$ & $+7$ & $+1$ & $+1$ & $+5$ & $+6$ \\
    \bottomrule
  \end{tabular}
\end{table}

\begin{figure}[t]
    \centering
    \includegraphics[width=\linewidth]{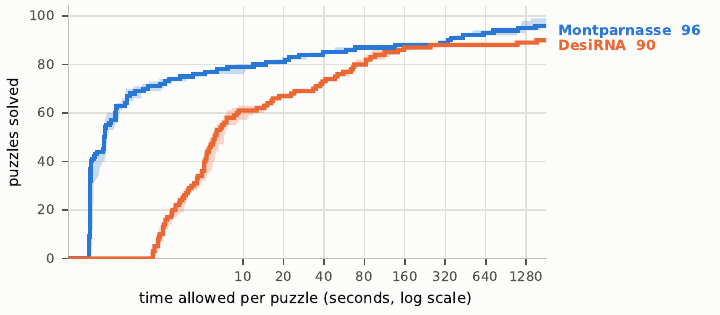}
    \caption{Eterna100 V1 puzzles solved as a function of the time allowed per
    puzzle, on the logarithmic scale of Table~\ref{tab:eterna100v1} and under
    the conditions of its caption. Each curve is the pointwise median of
    five passes, shaded between their minimum and maximum; they end at 95, 96,
    96, 97 and 99 for Montparnasse and at 89, 89, 90, 91 and 91 for DesiRNA.
    The clock is external and includes the startup of the fleet and the
    replicas.}
    \label{fig:courbe}
\end{figure}

\subsection{Experimental Setup}

All experiments are run on a single 256-core server with AMD EPYC-Rome
processors and 256~GB of RAM. Montparnasse uses 100 worker processes and
DesiRNA 100 replicas, and the hundred Eterna100~V1 puzzles are processed
sequentially, one puzzle at a time, so that each puzzle has the whole machine.

Both solvers fold through the same library. The energy model is Turner 1999 as
implemented by ViennaRNA 1.8.5, with dangling ends treated in the
\texttt{dangles=1} mode, and DesiRNA is made to call that same shared library
instead of its own folding back-end, so that the two searches optimize exactly
the same objective. This matters more than it may seem: a solution is a
sequence whose minimum free energy structure equals the target \emph{under a
given model}, and two models disagree on enough sequences that a comparison
across models measures the models as much as the searches. DesiRNA is run at
commit \texttt{bdb4908} with 100 replicas and an exchange every 100 steps.

The budget is 30 minutes of wall-clock time per puzzle for both solvers. The
clock is the external one: it starts before the fleet or the replicas are
launched and includes their startup, which is why no solving time is below
0.7~s for Montparnasse and 1.9~s for DesiRNA. A puzzle counts as solved only
if the sequence that is returned is folded again and verified to be at
base-pair distance 0 from the target.

\subsection{Results on Eterna100 V1}

Table~\ref{tab:eterna100v1} and Figure~\ref{fig:courbe} compare Montparnasse
and DesiRNA on Eterna100~V1 under this setup. Every number here is a median over
five passes, on each side. Montparnasse is ahead at every time limit of the
table, and at the end of the budget its median is 96 puzzles against 90.

The gap is widest at the short limits, where it is a gap in the quality of the
rollout rather than in the search: Montparnasse solves 55 of the hundred
puzzles within the first second and 79 within ten seconds, against 61 for
DesiRNA at ten seconds and none at all before 1.9~s. Those first fifty-five are
solved by the prior of Section~\ref{sec:prior} almost on its own, before any
adaptation of the policy has had time to matter. The gap then closes
entirely: the two medians are level at 88 between 250 and 290 seconds, and it
widens again beyond 320~s as the fleet pays on the hard targets.

Per puzzle the median is over the five times, a failing pass counting as
infinite. Ninety-seven have a finite median for Montparnasse and ninety
for DesiRNA; the 96 of Table~\ref{tab:eterna100v1} is the median of the five
counts, a different statistic, which here agrees for DesiRNA and not for
Montparnasse. On the 90 both solve, Montparnasse is faster on 88, with a
median solving time of 0.9~s against 5.6~s and a sum of 2{,}140~s against
4{,}645~s. It solves seven that DesiRNA does not, 73, 78, 85, 90, 91, 95
and 97, and DesiRNA none that it does not. A whole pass costs Montparnasse
a median of 12{,}784~s of machine time, the unsolved puzzles included, against
22{,}795~s.

\begin{table}[t]
  \centering\scriptsize\setlength{\tabcolsep}{2.6pt}
  \caption{The same comparison in the energy model earlier work on Eterna100
  has published in: ViennaRNA 2.6.4 with the Turner 1999 parameters and
  therefore with \texttt{dangles=2}. Montparnasse keeps its 100 workers and
  DesiRNA its 100 replicas, and the budget is one day per puzzle; only the
  energy model comes from earlier work. Each row is one pass, and the two
  passes were run one after the other on the same machine. Both reach the
  hundred, and the total wall-clock time to solve them sequentially is
  7,273~s for Montparnasse against 91,839~s for DesiRNA. These numbers are
  not comparable to those of
  Table~\ref{tab:eterna100v1}: see the text.}
  \label{tab:dangles2}
  \begin{tabular}{lrrrrrrrrrrrrrr}
    \toprule
    & \multicolumn{14}{c}{Puzzles solved within a time limit (seconds)} \\
    \cmidrule(l){2-15}
    & 10 & 20 & 40 & 80 & 160 & 320 & 640 & 1{,}280 & 2{,}560 & 5{,}120 &
      10{,}240 & 20{,}480 & 40{,}960 & 81{,}920 \\
    \midrule
    DesiRNA      & 60 & 67 & 75 & 85 & \textbf{93} & 93 & 93 & 94 & 96 & 97 &
                   97 & 99 & 99 & 100 \\
    Montparnasse & \textbf{80} & \textbf{81} & \textbf{87} & \textbf{89} & 92 &
                   \textbf{94} & \textbf{96} & \textbf{99} & \textbf{100} &
                   \textbf{100} & \textbf{100} & \textbf{100} & \textbf{100} &
                   \textbf{100} \\
    \bottomrule
  \end{tabular}
\end{table}

\subsection{The Energy Model Used in Earlier Work}

Table~\ref{tab:dangles2} gives the same comparison in the energy model
GREED-RNA and DesiRNA publish their results in: ViennaRNA 2.6.4 with the
Turner 1999 parameters. Only the model comes from that work: the one-day
budget and the hundred workers are our own choice, not a convention of
Eterna100. We report the model because the published numbers
are in it, and because both solvers there reach full coverage.

They do not reach it at the same price. Montparnasse solves the hundred
puzzles in 7{,}273~s of wall-clock time in all, two hours for the whole
benchmark, against 91{,}839~s for DesiRNA, twelve and a half times more. It
is the faster on 95 of the hundred, with a median solving time of 0.9~s
against 6.4~s, and its slowest target, puzzle 90, takes 2{,}064~s where
DesiRNA spends 48{,}047~s on its own slowest, puzzle 97. DesiRNA is
nevertheless ahead at one limit of the table, 160~s, where it has 93 against
92. Two things separate this configuration from the one of
Table~\ref{tab:eterna100v1} besides the budget: the word rule of
Section~\ref{sec:fleet} is not applied here, and the fleet restart of the
same section is on, with its floor at 1{,}800~s.

It is not, however, the right constraint. ViennaRNA 2.6.4 treats dangling ends
in its \texttt{dangles=2} mode, in which a dangling end contributes its
stabilizing energy to every adjacent helix at once, whether or not it could
physically stack on both. That is not the configuration Eterna scores its
puzzles in, and the substitution is not neutral: the two models disagree about which sequences fold into a
target, and the disagreement runs in the easy direction. Puzzle 100
shows it at its sharpest. Under \texttt{dangles=2} it is among the last of
the hundred to fall, but it falls in every long run we have given it, in
890~s, 5{,}131~s and 18{,}572~s. Under \texttt{dangles=1} one thirty-minute
pass in five reaches it and no longer run ever has, and puzzle 99 needs a
dedicated run of 5{,}028~s. Same machine, and searches that differ only by
the word rule and the restart named above, neither of which changes which
sequences fold into the target; the treatment of dangling ends does. The two
tables therefore measure
different problems and must not be read against each other. Everything else
in this paper is measured under ViennaRNA 1.8.5 with \texttt{dangles=1}.

\subsection{The Two Hard Puzzles}

Puzzles 99 and 100 are the two targets on which the fleet mechanisms of
Section~\ref{sec:fleet} were designed. They are hard in two different ways,
which the beam makes observable, and only one of them is reached inside the
30-minute budget.

Puzzle 100 is the one whose failure is legible. The fleet reaches base-pair
distance 1 or 2 and the whole beam then fails on the same handful of pairs:
the dominant fault is an extra pair of the same shape as the target's own
helices, formed between two distant units of the molecule, and no local repair
removes it without breaking something else. It is nevertheless solved in one
of the five passes, after 389~s on a target of 382 nucleotides, and how it is solved is the
observation. Long runs on this target fail: the two
one-hour passes we ran under this model leave it unsolved, and so do the
dedicated runs we gave it before them. The run that gets it is a short one.
Under \texttt{dangles=2}, where the long runs do succeed, the restart tells
the same story. Three of them reach the target: 890~s with 100 workers and
5{,}131~s with 50, both restarting the fleet after 1{,}800~s without
progress, against 18{,}572~s with 100 workers and no restart. The slowest is
the one that searches without interruption, and it is not the one with the
smaller fleet. On this target the solving time behaves like a heavy-tailed
distribution, in which restarting early beats searching long, what the
fleet restart of Section~\ref{sec:fleet} is for. Each is a single run, and we
do not extend the conclusion to the rest of the benchmark.

Puzzle 99 is a matter of time, and it is the one the budget does not reach.
Over its plateau at base-pair distance 2, the intersection of the fault lists
of the hundred ranks of the beam is empty at every dump, the hundred ranks
carry a hundred distinct families, and from one dump to the next the set of
families is renewed by between 13 and 44\%. There is no common lock to attack:
the fleet fails in a hundred different places at once rather than on one wall.
What it needs is a longer budget, and given one, a dedicated run of the
same configuration, with no time limit, Montparnasse solves it under this
same model, in 5{,}028~s. Every one of the hundred targets is therefore
solved under ViennaRNA~1.8.5 with \texttt{dangles=1}, and the hundred
solutions are listed in the appendix, each accepted by the Eterna validator
folding with the stock 1.8.5 binary. What is not claimed is that one configuration reaches
all hundred within one budget: the five passes of
Table~\ref{tab:eterna100v1} share one, and none of them does.

\section{Conclusion}

Montparnasse is a Monte Carlo search framework for Inverse RNA Folding that
combines Generalized Nested Rollout Policy Adaptation with five components: a
structural prior on the policy derived from the energy model and the topology
of the target, slow and long adaptation at level 1 with a patience
stopping criterion, a lexicographic multicriteria evaluation,
a local repair of the pairs the predicted folding gets wrong, and a
fleet of parallel workers coupled through a shared beam.

On the Eterna100 V1 benchmark, with both solvers folding through the same
ViennaRNA 1.8.5 library under the Turner 1999 model and a budget of 30 minutes
per puzzle, Montparnasse with 100 workers solves a median of 96 of the 100
puzzles while DesiRNA with 100 replicas solves a median of 90, five passes on
each side. Montparnasse is ahead at every time limit of
Table~\ref{tab:eterna100v1}, it is faster on 88 of the 90 puzzles both solve,
and it solves seven puzzles DesiRNA does not and none the other way. Puzzle 99, which
no pass reaches, is solved under the same model with a longer budget, so every
Eterna100~V1 target is reached. In the
\texttt{dangles=2} setting of Table~\ref{tab:dangles2}, in which GREED-RNA
and DesiRNA publish their results, both solvers reach the full hundred
within a budget of one day per puzzle, Montparnasse in 7{,}273~s of
wall-clock time against 91{,}839~s; but that setting is not the one Eterna
scores its puzzles in and it makes them easier, so we do not take it as the
result.

The component we would single out is the admission rule of the beam. Measuring
the diversity of a population of sequences on the sequences themselves
is natural and wrong: seven front sequences a distance rule
held to be diverse carried one and the same list of faults, and a beam of a
hundred distinct fault families was one design with a hundred one-helix
mutants around it. What
a parallel search needs spread out is its \emph{failures}, and a failure is
observable: the list of pairs the predicted folding gets wrong.
Both rules of Section~\ref{sec:fleet} follow from it, and both were
validated by a control run in which they alone changed.

The components introduced in Montparnasse are not all specific to RNA design.
The prior is, and so is the local repair; the slow and long adaptation at
level 1, and the beam whose diversity is stated on the faults rather than on
the candidates, apply to GNRPA in general and transfer to any combinatorial
optimization problem amenable to it, provided its failures can be named. Future work
will apply them to other domains where GNRPA has been used, and seek
one configuration that solves all of Eterna100~V1 in a single budget, which
none here does.

\bibliographystyle{splncs04}
\bibliography{Montparnasse}

\begin{thebibliography}{10}
\providecommand{\url}[1]{\texttt{#1}}
\providecommand{\urlprefix}{URL }
\providecommand{\doi}[1]{https://doi.org/#1}

\bibitem{eterna}
Anderson-Lee, J., Fisker, E., Kosaraju, V., Wu, M., Kong, J., Lee, J., Lee, M.,
  Zada, M., Treuille, A., Das, R.: Principles for predicting rna secondary
  structure design difficulty. Journal of Molecular Biology  \textbf{428}(5,
  Part A),  748--757 (2016)

\bibitem{andronescu2004new}
Andronescu, M., Fejes, A.P., Hutter, F., Hoos, H.H., Condon, A.: A new
  algorithm for {RNA} secondary structure design. Journal of molecular biology
  \textbf{336}(3),  607--624 (2004)

\bibitem{bonnet2020designing}
Bonnet, {\'E}., Rza{\.z}ewski, P., Sikora, F.: Designing {RNA} secondary
  structures is hard. Journal of Computational Biology  \textbf{27}(3),
  302--316 (2020)

\bibitem{busch2006info}
Busch, A., Backofen, R.: {INFO-RNA}—a fast approach to inverse {RNA} folding.
  Bioinformatics  \textbf{22}(15),  1823--1831 (2006)

\bibitem{CazenaveIJCAI09}
Cazenave, T.: {Nested Monte-Carlo Search}. In: Boutilier, C. (ed.) IJCAI. pp.
  456--461 (2009)

\bibitem{Cazenave2020GNRPA}
Cazenave, T.: Generalized nested rollout policy adaptation. In: Monte Search at
  IJCAI (2020)

\bibitem{cazenave2024generalized}
Cazenave, T.: Generalized nested rollout policy adaptation with limited
  repetitions. European Workshop on Reinforcement Learning  (2024)

\bibitem{cazenave2024learning}
Cazenave, T.: Learning a prior for monte carlo search by replaying solutions to
  combinatorial problems. In: International Conference on Parallel Problem
  Solving from Nature. pp. 85--99. Springer (2024)

\bibitem{Cazenave2020Inverse}
Cazenave, T., Fournier, T.: Monte {C}arlo inverse folding. In: Monte Search at
  IJCAI (2020)

\bibitem{cazenave2024monte}
Cazenave, T., Touzani, H.: {Monte Carlo} inverse {RNA} folding. In: RNA Design:
  Methods and Protocols, pp. 205--215. Springer (2024)

\bibitem{churkin2024rna}
Churkin, A., Barash, D.: RNA Design. Springer (2024)

\bibitem{hofacker1994fast}
Hofacker, I.L., Fontana, W., Stadler, P.F., Bonhoeffer, L.S., Tacker, M.,
  Schuster, P., et~al.: Fast folding and comparison of {RNA} secondary
  structures. Monatshefte fur chemie  \textbf{125},  167--167 (1994)

\bibitem{lozano2024simple}
Lozano-Garc{\'\i}a, N., Rubio-Largo, {\'A}., Granado-Criado, J.M.: A simple yet
  effective greedy evolutionary strategy for {RNA} design. IEEE Transactions on
  Evolutionary Computation  (2024)

\bibitem{merleau2022arnaque}
Merleau, N.S., Smerlak, M.: {aRNAque}: an evolutionary algorithm for inverse
  pseudoknotted rna folding inspired by l{\'e}vy flights. BMC bioinformatics
  \textbf{23}(1), ~335 (2022)

\bibitem{portela2018unexpectedly}
Portela, F.: An unexpectedly effective {M}onte {C}arlo technique for the {RNA}
  inverse folding problem. BioRxiv p. 345587 (2018)

\bibitem{Reese2005}
Reese, C.B.: Oligo- and poly-nucleotides: 50 years of chemical synthesis. Org.
  Biomol. Chem.  \textbf{3},  3851--3868 (2005)

\bibitem{Rosin2011}
Rosin, C.D.: Nested rollout policy adaptation for {Monte Carlo Tree Search}.
  In: IJCAI. pp. 649--654 (2011)

\bibitem{taneda2015multi}
Taneda, A.: Multi-objective optimization for {RNA} design with multiple target
  secondary structures. BMC bioinformatics  \textbf{16},  1--20 (2015)

\bibitem{wirecki2025desirna}
Wirecki, T.K., Lach, G., Badepally, N.G., Moafinejad, S.N., Jaryani, F.,
  Klaudel, G., Nec, K., Baulin, E.F., Bujnicki, J.M.: Desirna: structure-based
  design of rna sequences with a replica exchange monte carlo approach. Nucleic
  Acids Research  \textbf{53}(2),  gkae1306 (2025)

\bibitem{yang2017rna}
Yang, X., Yoshizoe, K., Taneda, A., Tsuda, K.: Rna inverse folding using monte
  carlo tree search. BMC bioinformatics  \textbf{18},  1--12 (2017)

\end{thebibliography}

\section*{Appendix: the solutions}

The 100 sequences below are the solutions Montparnasse found, each with the time at which it was found. They fold into their target under ViennaRNA~1.8.5 with the Turner~1999 parameters and \texttt{dangles=1}, which is the model Eterna100~V1 was built and validated in. All but one come from the first of the five 30-minute passes of Table~\ref{tab:eterna100v1}; puzzle 99, marked $^{\dagger}$, was found in a dedicated run with no time limit, as described in Section~\ref{sec:experiments}. Each one was refolded and checked against its target after the run. Each one was then checked again by \texttt{check\_design.py} of the Eterna validator, which folds with the stock ViennaRNA~1.8.5 \texttt{RNAfold} binary rather than with the library the search itself uses, under the published benchmark condition that the minimum free energy structure equal the target: 100 of 100 pass.

\begingroup\footnotesize\setlength{\parindent}{0pt}\raggedright
\textbf{1}\quad 0.76~s, 16~nt\\
\texttt{GGGCGAAAAA\allowbreak GCGCCC}\par\smallskip
\textbf{2}\quad 0.74~s, 116~nt\\
\texttt{GGGGCCGGAA\allowbreak ACCCGGAAGA\allowbreak AAAAAGGGCG\allowbreak GCGCGGUUCG\allowbreak CCGGCGGGCG\allowbreak GCUCGGCGAG\allowbreak GGGGGGAAAA\allowbreak ACCCCCAGCC\allowbreak GGGACCGCCC\allowbreak GCCGGCGGUC\allowbreak CGCGCCGAGC\allowbreak GGCCCC}\par\smallskip
\textbf{3}\quad 0.75~s, 36~nt\\
\texttt{CGCGCGAGGC\allowbreak CGAGUGGCCC\allowbreak GCAGCGAAAA\allowbreak AAAAAC}\par\smallskip
\textbf{4}\quad 1.37~s, 192~nt\\
\texttt{GCCGGCGCGC\allowbreak GGGGACCGGG\allowbreak GAAAAAAAAA\allowbreak AAGCGGCGGC\allowbreak GGGCAAGAAC\allowbreak CCCCGGGAAA\allowbreak ACGCCCGCGC\allowbreak CCCCGAAAAA\allowbreak AAGGGGGCGC\allowbreak GGAAAAGCGC\allowbreak CCGGAGAAGG\allowbreak GGCCAAAACG\allowbreak CCGAAAGGGG\allowbreak CGGAAAAAAC\allowbreak GCCCCCCGCG\allowbreak AAAAAAAAAA\allowbreak ACCCCGGGCC\allowbreak CCGGAAGCGC\allowbreak GCAAACGGCA\allowbreak AA}\par\smallskip
\textbf{5}\quad 0.75~s, 101~nt\\
\texttt{AAGAGUGGCC\allowbreak CCGGAGACGG\allowbreak GGCUGCGCGA\allowbreak AAAACGCCAA\allowbreak GGCACGGCCG\allowbreak GCGGAGACGC\allowbreak CGGAAAGCCG\allowbreak GCCGGCGAAG\allowbreak GGGCCGCGAA\allowbreak AGCGGCCCCA\allowbreak A}\par\smallskip
\textbf{6}\quad 1.15~s, 379~nt\\
\texttt{CCCGGCCCCG\allowbreak GAGAAAAAAC\allowbreak CGGAGGGAAA\allowbreak AAGAAAAAAA\allowbreak ACCAAAACGG\allowbreak GGAGAAAAAC\allowbreak CCCGGGCCGG\allowbreak GAAAAGCCGG\allowbreak UCAACCCGGA\allowbreak CAAAAACGGG\allowbreak GGCAAAAACA\allowbreak AAGAGAGCCA\allowbreak AAGCGCCGGA\allowbreak AAGAACGGCG\allowbreak GGCCGGCAAA\allowbreak ACCACCGGAA\allowbreak GGCCGAAAAA\allowbreak AAGGCCCCGC\allowbreak CGAAAAAAAC\allowbreak GGCGGAAAAC\allowbreak GGGCGAAAGA\allowbreak AGCCCGCCGG\allowbreak UGGGAAAGGG\allowbreak CGUCAAGGCG\allowbreak GAAAAAAACG\allowbreak CCCCGGCGGA\allowbreak AGAAACGCCG\allowbreak GAAAGCGCGC\allowbreak AGAAAAAGCG\allowbreak CGGACGCCCA\allowbreak AAACCCGGGC\allowbreak AACGGGGAAA\allowbreak AAAACCCGGC\allowbreak GGGGGAAAAA\allowbreak ACCCCGCAAG\allowbreak AGCGCCAAAA\allowbreak AAAGGCGCGC\allowbreak CCGGGAAGA}\par\smallskip
\textbf{7}\quad 0.77~s, 284~nt\\
\texttt{AACAAGGGCC\allowbreak CGCAAACCUG\allowbreak GCAGCCAACG\allowbreak CCAAGAAAAA\allowbreak AAGGCGAACG\allowbreak CGGAGAAAAA\allowbreak AACGCGAACU\allowbreak CCGAAAAAAA\allowbreak AAGGAGAAGG\allowbreak CUGCCGGGAA\allowbreak GCCCCCCGCG\allowbreak GAAGCGCAAA\allowbreak AAAAAAAGCG\allowbreak CAACGGGAAA\allowbreak GAAAAGACCC\allowbreak GAAGGCGGAA\allowbreak AAAAAAACGC\allowbreak CAACCGCGGG\allowbreak GGGAAAUGCC\allowbreak GCGGCCAACG\allowbreak CGGAAAAAAA\allowbreak AACGCGAGGC\allowbreak GCGGAAAAAA\allowbreak AAGCGCAACG\allowbreak CCGAAAACAA\allowbreak GAGGCGAAGG\allowbreak CCGCGGCAAA\allowbreak AGCGGGUCCA\allowbreak AAAA}\par\smallskip
\textbf{8}\quad 0.76~s, 12~nt\\
\texttt{CGGCAAAGCC\allowbreak GA}\par\smallskip
\textbf{9}\quad 2.67~s, 258~nt\\
\texttt{GGGAGGCCCG\allowbreak GAAACCGGAA\allowbreak GCCCCCAAAC\allowbreak CGACGCCCGC\allowbreak AAAGCGGAAG\allowbreak CGCGGAAACC\allowbreak GACGGCCGGA\allowbreak AACCGGAACC\allowbreak GCGGGAACGG\allowbreak AGCGCGGGAG\allowbreak ACCCGAACGC\allowbreak CCGAAACCCA\allowbreak CCGGCCCAAG\allowbreak GGGCAACGGG\allowbreak GGAAAGGCAG\allowbreak GCCCGGAAAC\allowbreak CGGAAGCCGC\allowbreak CAAAGCGACG\allowbreak GCCGCGAAGC\allowbreak GGAACCGCGC\allowbreak AAACGCACGG\allowbreak CCGCAAAGCG\allowbreak GAACCGGCGG\allowbreak AACGCACGGC\allowbreak GCGAAACGCG\allowbreak AACCGGCG}\par\smallskip
\textbf{10}\quad 0.75~s, 45~nt\\
\texttt{AAAAAAAAAA\allowbreak GGUGGAAACA\allowbreak CCCCCGGGAA\allowbreak CGGGCCGCAA\allowbreak AGCGG}\par\smallskip
\textbf{11}\quad 0.76~s, 36~nt\\
\texttt{GCCGUGAGCG\allowbreak CGGGGGGGAC\allowbreak CCUGACGACA\allowbreak CGCGGC}\par\smallskip
\textbf{12}\quad 0.84~s, 213~nt\\
\texttt{GGCGGAAACG\allowbreak CCAAAGGCAC\allowbreak GGAACCGGCG\allowbreak GGGAAAAACC\allowbreak GCCGGCCGUG\allowbreak CCAAGGCCGU\allowbreak GCCGAGAAAG\allowbreak GGCGGGUGGG\allowbreak GCCGGAAAAA\allowbreak GCGGCCCCCC\allowbreak CGCCCAAACG\allowbreak GGCGCGGGGG\allowbreak GCGAAAAAAG\allowbreak CCCCCCAGCG\allowbreak CCCGAAAAAA\allowbreak GGCACGGCAA\allowbreak AGGGCCGCAC\allowbreak CCGCCAAGAA\allowbreak AGGGCGGGUA\allowbreak AGCGGCCCAA\allowbreak ACUCGGAUCC\allowbreak GAG}\par\smallskip
\textbf{13}\quad 0.77~s, 67~nt\\
\texttt{CGCGCCCGCG\allowbreak CGGCGGCGCG\allowbreak CCCGGCAAAG\allowbreak CCGGGGAGAC\allowbreak GCGCCGUAAA\allowbreak CCGCGCAAAA\allowbreak GGGCGCG}\par\smallskip
\textbf{14}\quad 0.97~s, 98~nt\\
\texttt{AAAAACCGAA\allowbreak GGGAGACCAG\allowbreak AGCGAACGCA\allowbreak AAGGGAAACC\allowbreak AAAGGGAGAC\allowbreak CAGAGGGAGA\allowbreak CCAAACCGAG\allowbreak AGGAAAGGAA\allowbreak AGAAGGAAAA\allowbreak AAGAAAAA}\par\smallskip
\textbf{15}\quad 0.77~s, 30~nt\\
\texttt{GGCGCGAAAA\allowbreak GCAAGGGAAA\allowbreak AAAAACCGCC}\par\smallskip
\textbf{16}\quad 0.75~s, 105~nt\\
\texttt{AGAAGAAAAA\allowbreak AACGGAAACG\allowbreak CAAGCAAAGG\allowbreak GAGACCAACC\allowbreak AAACGGAGCC\allowbreak GGAGGGAAGC\allowbreak CAAAGCAAAG\allowbreak CAACCGAACG\allowbreak GAAAAAACUG\allowbreak AAGAGAAACA\allowbreak AGAAA}\par\smallskip
\textbf{17}\quad 0.74~s, 105~nt\\
\texttt{AAGAAGGAAA\allowbreak ACGGAAACGA\allowbreak AAAGGGGAAC\allowbreak CAAGAGCGAU\allowbreak AGCAAAAGGG\allowbreak AGACCAAAAG\allowbreak GGGGACCAAA\allowbreak AGGGAGACCA\allowbreak AAACCAAAAA\allowbreak AAAGAGAAAA\allowbreak AAAAA}\par\smallskip
\textbf{18}\quad 0.85~s, 67~nt\\
\texttt{GCGCCCAAAG\allowbreak GGCCGAAAGG\allowbreak GAAACGGGCG\allowbreak AAAAAGAAAA\allowbreak AAGCCCGCAA\allowbreak GCCCAAGGCG\allowbreak GGGGCGC}\par\smallskip
\textbf{19}\quad 0.77~s, 104~nt\\
\texttt{AAAAACCGGA\allowbreak CGGGCGAGAG\allowbreak CACCGAACCA\allowbreak GCGGAAACGC\allowbreak GGUGCCCGCG\allowbreak GAGACGACGG\allowbreak GAGGCUGGAG\allowbreak GCAGACCCCG\allowbreak GAAAAGAAAA\allowbreak AAAAAAAAAG\allowbreak AAGG}\par\smallskip
\textbf{20}\quad 0.75~s, 42~nt\\
\texttt{CGCCGGGAGA\allowbreak GAGAGGGUGC\allowbreak CGAGAGGCAC\allowbreak CCCCCCCCGG\allowbreak CG}\par\smallskip
\textbf{21}\quad 0.96~s, 105~nt\\
\texttt{AAAAAGGCCC\allowbreak CGCGAAAGCG\allowbreak GCGGGACCCC\allowbreak GCAAGGGGCG\allowbreak CCGGACACGG\allowbreak CGACCCGGAA\allowbreak ACGGGCCCGG\allowbreak CGGGAGACCG\allowbreak CGGCCAAGAA\allowbreak CAAAAAAAAA\allowbreak AAGAA}\par\smallskip
\textbf{22}\quad 0.78~s, 400~nt\\
\texttt{AGAAAAAAAA\allowbreak AAAAAAAAAA\allowbreak AAAAAAAAAA\allowbreak AAAAAAAAAA\allowbreak AAAAAAAAAA\allowbreak AAAAAAAAAA\allowbreak AAGAAAAAAA\allowbreak AAAAAAAGAA\allowbreak AAAAAAGAAG\allowbreak GAAAGAAAAA\allowbreak AAAAAAAAAC\allowbreak AAGAAAAAAA\allowbreak AAAAAAAAAA\allowbreak AAAAAGAAAA\allowbreak AAAACAAAAA\allowbreak AAGGAAAAAA\allowbreak AAAGAGGAAA\allowbreak AAAAAAAAAA\allowbreak GAAAAAAAAA\allowbreak AACAAAAAAA\allowbreak AAAAAAAAAA\allowbreak GAGAAAAAAA\allowbreak AAAAAAAAAA\allowbreak AAAAAAAAAA\allowbreak AAAAAAAAAA\allowbreak AAAAAGAAAA\allowbreak AAGAAGAACA\allowbreak AGGAAAAGAA\allowbreak AGGAAAAAAA\allowbreak GAACAAAAAA\allowbreak AAAAAAAAAA\allowbreak AAAAAAGAAA\allowbreak AAAAAAAAAA\allowbreak AAAAAAAAAA\allowbreak AAAAAAAAAG\allowbreak AAAAAAAAAA\allowbreak AAAAAAGAAA\allowbreak AAGAAAAAGA\allowbreak AAAGAAAAAA\allowbreak GAAAAGAAAA}\par\smallskip
\textbf{23}\quad 0.76~s, 17~nt\\
\texttt{GGGAAACCAG\allowbreak GGAAACC}\par\smallskip
\textbf{24}\quad 0.77~s, 104~nt\\
\texttt{ACAAACGGGG\allowbreak CAGGCCGGAA\allowbreak AAAAAACCGG\allowbreak CAAGCCAUCC\allowbreak ACCGCCGAGA\allowbreak AGGCGGAGCC\allowbreak GGAAAUGGCA\allowbreak GGGGGCAGGC\allowbreak CCGGAAAAGG\allowbreak AAUAAGGAAA\allowbreak AGAA}\par\smallskip
\textbf{25}\quad 0.76~s, 62~nt\\
\texttt{CCCAGAAACG\allowbreak GAAAAGCGAA\allowbreak AAAACCGAAA\allowbreak AGGCGAAAAA\allowbreak AAACGCGCGG\allowbreak GUCGACCGAG\allowbreak GG}\par\smallskip
\textbf{26}\quad 0.76~s, 26~nt\\
\texttt{AGGGGGCGCG\allowbreak AAAAACGACG\allowbreak CCCCAA}\par\smallskip
\textbf{27}\quad 0.73~s, 57~nt\\
\texttt{GAAAACGCCC\allowbreak AGAGAGGAGA\allowbreak CCGGGGCGGG\allowbreak AAACCGCCCC\allowbreak GGCCCCCGGG\allowbreak CGAAAAG}\par\smallskip
\textbf{28}\quad 1.18~s, 378~nt\\
\texttt{AUACCGCAAG\allowbreak GCGGAAAAAC\allowbreak CGCCAACGAG\allowbreak GAAACCUCGA\allowbreak ACCGGGAAAA\allowbreak ACCCGGAAGG\allowbreak CCCAAAGGGC\allowbreak CAAGCUGGAG\allowbreak AAACCAGCAA\allowbreak GGGCCGGAAA\allowbreak GGCCCAGCCG\allowbreak GGGGCCCGAG\allowbreak CGCUCCGGGC\allowbreak GGGCGAAAGA\allowbreak GCCCGACGGC\allowbreak GGAAGAAAAA\allowbreak AAGGGGCGAA\allowbreak AAAAGCAAAA\allowbreak AAAAAAAGAA\allowbreak AGAAAAAGAA\allowbreak AAAAAAAAAC\allowbreak GCCAAAAAGA\allowbreak AAAGAGCGCC\allowbreak GGCCGCGAAA\allowbreak GCGGCACCCG\allowbreak GGGCGCUCGG\allowbreak GUCCCCGGCA\allowbreak ACGCCCGAAG\allowbreak AGGGCGAACC\allowbreak CCCGAAGAGG\allowbreak GGGAAGCCGG\allowbreak AGAAACCGGC\allowbreak GACCGCGGGG\allowbreak AACGCGGGAC\allowbreak GCCGGAGACG\allowbreak GCGAACCCGC\allowbreak GAAAAGCGGG\allowbreak AGCGGAAA}\par\smallskip
\textbf{29}\quad 1.57~s, 105~nt\\
\texttt{AAAAGGCCCC\allowbreak CGGGUCGGCG\allowbreak AGAGCCGUUC\allowbreak CGGAGGCGUG\allowbreak CGAAGUGCGC\allowbreak CAGGGGAAGA\allowbreak GCGCAAAGGG\allowbreak CGGCCAGCCC\allowbreak CGGGCAAAAA\allowbreak AAGGAAAAAA\allowbreak GAAAA}\par\smallskip
\textbf{30}\quad 0.80~s, 31~nt\\
\texttt{AGGCGGCCGC\allowbreak CCAAAGGGCG\allowbreak AAAAGCCGCC\allowbreak A}\par\smallskip
\textbf{31}\quad 0.77~s, 102~nt\\
\texttt{AGAAAGCGGA\allowbreak CGGCGCACGC\allowbreak AGGCGCAGGC\allowbreak CAGGCGAGCA\allowbreak CGAGGAGGAC\allowbreak GACGGAAAGC\allowbreak GCGCCCCCGG\allowbreak CCGCCGGCCG\allowbreak CGCCGCGGCG\allowbreak CCGCCGCAAA\allowbreak AG}\par\smallskip
\textbf{32}\quad 1.37~s, 122~nt\\
\texttt{AAAAAAAAAA\allowbreak AAAAGAAAGA\allowbreak AAAAAAAACG\allowbreak CCUAGAGAAG\allowbreak GCCGAAGGCG\allowbreak CAGGCGGGAG\allowbreak GCCCGCGGCA\allowbreak AAAAGCACGA\allowbreak GGAGCGGCCA\allowbreak CGAAAGCAGC\allowbreak AAAAGGGACC\allowbreak GAAAAAAGAG\allowbreak CA}\par\smallskip
\textbf{33}\quad 0.76~s, 45~nt\\
\texttt{GAAAAAAGGG\allowbreak AGGCGGACGG\allowbreak GCAGGAAACG\allowbreak GCGCGGCGGG\allowbreak CGCGC}\par\smallskip
\textbf{34}\quad 0.83~s, 96~nt\\
\texttt{AAAGACCGCA\allowbreak CCGGGGAUAC\allowbreak CCGGACCCGA\allowbreak GAGGGAGCAG\allowbreak GAAAGCCCCA\allowbreak CCGCAGACGG\allowbreak AGCCGAUAGG\allowbreak CAGGGGAAAA\allowbreak AAAAAAGAAA\allowbreak AAAAAA}\par\smallskip
\textbf{35}\quad 1.00~s, 123~nt\\
\texttt{AGCGAGCCAA\allowbreak ACCCGCGGGA\allowbreak ACCGAGGCGA\allowbreak GAGCCGGGGG\allowbreak AACGCAGCCG\allowbreak AGAGGCGCCG\allowbreak AAACGGAGCG\allowbreak AAAGCCAGCC\allowbreak GAGAGGCAGG\allowbreak CAAAGCCAGG\allowbreak CGAAGGCAGC\allowbreak GAGACGCACG\allowbreak CAA}\par\smallskip
\textbf{36}\quad 0.76~s, 151~nt\\
\texttt{AAAAAGGGCC\allowbreak GCAAAGGGGG\allowbreak GAGGCGAAAC\allowbreak GGGAAAAAAG\allowbreak AACGCCGAAA\allowbreak AAAAGAAAAC\allowbreak GGCGCAACCC\allowbreak GAAAACCGCG\allowbreak CAAAAAAAAG\allowbreak CGGAAACGCC\allowbreak AAGGCGAAGA\allowbreak AAGACGCAGA\allowbreak CCCCAGGCGA\allowbreak AGAAGCAAAG\allowbreak CGGCCCAAAA\allowbreak A}\par\smallskip
\textbf{37}\quad 1.36~s, 103~nt\\
\texttt{AAAAGCCAGG\allowbreak UGCCGGGACG\allowbreak GGGGACCGCG\allowbreak CGAAAGCGCC\allowbreak CAAAAACGUC\allowbreak CGGAAACGGG\allowbreak GCGAGAGCCA\allowbreak GCGGGUGCCA\allowbreak GGAAGAAGAA\allowbreak GAGAAAAAAA\allowbreak AAA}\par\smallskip
\textbf{38}\quad 0.97~s, 316~nt\\
\texttt{AAAGGGAAAC\allowbreak CAAAACCCGG\allowbreak CCUGGAGCAA\allowbreak AAACGGGCCG\allowbreak GGAAAAGCGG\allowbreak AAAAACCGCA\allowbreak AAGAGCGGAG\allowbreak GCGCAAACGC\allowbreak GCGGAAGAGG\allowbreak CGCGAAAACC\allowbreak GAAACCGAAA\allowbreak GGAUAGACGG\allowbreak GAAACCGAAA\allowbreak ACGGGCCGAA\allowbreak AGGAGGCCCG\allowbreak GAAACGCGAG\allowbreak AGCGAAGAAG\allowbreak GGGGCAAAAC\allowbreak CCGAAGAAAG\allowbreak GGAAGCGGAA\allowbreak GAAAACGCAA\allowbreak AGCGGGAUAA\allowbreak AACCCGCAAA\allowbreak CGGGAGGAAC\allowbreak CCGAAGCCCC\allowbreak CCAAAAGAAA\allowbreak CGGAAACGAA\allowbreak GGAGGGAGAC\allowbreak CAAAAAACGG\allowbreak GGGAAAAAAA\allowbreak AACCCCGAAA\allowbreak ACACGG}\par\smallskip
\textbf{39}\quad 0.89~s, 174~nt\\
\texttt{ACGCCACCGC\allowbreak GGGCGCCCCA\allowbreak UCAGCGAGCC\allowbreak CCGACCGGGG\allowbreak AAAACCGCGC\allowbreak AAAAAAGCGC\allowbreak GGCCCCCGGA\allowbreak CGGGCAAAGG\allowbreak CGGGCAAAGC\allowbreak CCGCCGCCCG\allowbreak ACGGGGCAGC\allowbreak GGCGGCGGGC\allowbreak AAGGCCCGCC\allowbreak AAAGCCGCAC\allowbreak GCUGGGAAAG\allowbreak GGGCGGGAAC\allowbreak CGCGGAAAGG\allowbreak GCGU}\par\smallskip
\textbf{40}\quad 0.76~s, 38~nt\\
\texttt{GACGGCCCGG\allowbreak AAAAACCGGG\allowbreak GGCGAAAAGC\allowbreak CCGCCGAA}\par\smallskip
\textbf{41}\quad 0.77~s, 35~nt\\
\texttt{GCGAACGCAG\allowbreak GGAGACCCGG\allowbreak GAGACCAGGG\allowbreak AAACC}\par\smallskip
\textbf{42}\quad 0.98~s, 116~nt\\
\texttt{AAAGAGGCGG\allowbreak CGCCAAAAAG\allowbreak CGGAAACCGC\allowbreak AGGCGAAGGC\allowbreak GAACCCGCGA\allowbreak AACGCAAAGG\allowbreak GGCCCCGAAA\allowbreak CGGAAAGGCA\allowbreak CGCAAACGGA\allowbreak AAGCCGAGAG\allowbreak GCACCGACGC\allowbreak CAAAAG}\par\smallskip
\textbf{43}\quad 0.93~s, 85~nt\\
\texttt{AAAAACCUGG\allowbreak GAGGAGCCGA\allowbreak GCCGGCGGAG\allowbreak GGAGGCUGAG\allowbreak CGCCACCGAG\allowbreak GCGGGGCCGC\allowbreak CAGGAAAAAG\allowbreak AAGAAAAAAA\allowbreak AAAAA}\par\smallskip
\textbf{44}\quad 0.77~s, 63~nt\\
\texttt{AAAAAGGCAG\allowbreak CCCCGAAAGG\allowbreak AGGCGGCACG\allowbreak CAAACGGCCA\allowbreak GCCAAGAAAA\allowbreak AGAAUAAAAA\allowbreak GAA}\par\smallskip
\textbf{45}\quad 0.86~s, 63~nt\\
\texttt{GGCUCCCCGG\allowbreak AAACGGGUCC\allowbreak CCGAAAGGGG\allowbreak ACGCGCGAGA\allowbreak GCGCGGGGCG\allowbreak ACCGCCCGAG\allowbreak CCA}\par\smallskip
\textbf{46}\quad 1.20~s, 97~nt\\
\texttt{AAAAGCGGGG\allowbreak GCGGGGCGAA\allowbreak AGCCCCCGCG\allowbreak AAAGCGGCGC\allowbreak CGGGGUCCGG\allowbreak AGACGGAGCG\allowbreak GGAGACCGCC\allowbreak CCCCCCGAAA\allowbreak AAAAAAAAAA\allowbreak AAAAAAA}\par\smallskip
\textbf{47}\quad 0.75~s, 50~nt\\
\texttt{CGGGAAAAAA\allowbreak GGGCCAAAGG\allowbreak CAGGGAGACC\allowbreak AAAAAAAAAC\allowbreak CGAAAACCCG}\par\smallskip
\textbf{48}\quad 0.75~s, 102~nt\\
\texttt{AAAGAGGGGC\allowbreak CAGAAACUGG\allowbreak AGGCCGGGCA\allowbreak AAGCCCAACC\allowbreak GCAAAGCGGA\allowbreak ACGCGGAACG\allowbreak CGGGCCGGGA\allowbreak GAAACUCCCC\allowbreak CCAAAGAAAA\allowbreak AAAAAAAAAA\allowbreak AA}\par\smallskip
\textbf{49}\quad 2.00~s, 101~nt\\
\texttt{AAAGAGGGAC\allowbreak UGCGGCAGGA\allowbreak GCGGACCAAA\allowbreak GGCCAGCCCC\allowbreak GCCCGAAACG\allowbreak AGGCGAAGCC\allowbreak AAAGGCAAAU\allowbreak GCGGUGGACC\allowbreak CAAAAAAAAA\allowbreak AAAAAAAAAA\allowbreak A}\par\smallskip
\textbf{50}\quad 1.29~s, 105~nt\\
\texttt{AAAAAGGCGA\allowbreak GGCCCGAAAG\allowbreak GAAAAGCCAG\allowbreak GCGAAAGCCG\allowbreak CGACCAAGAG\allowbreak CCCAGCGGAG\allowbreak ACCAAGCAGG\allowbreak GGGGACCCAG\allowbreak GAAGCAAAAA\allowbreak AAAAAAAAAA\allowbreak AAAGA}\par\smallskip
\textbf{51}\quad 0.96~s, 337~nt\\
\texttt{GCGCCGAGCC\allowbreak GGCGAUAGCC\allowbreak GGCACCGGAG\allowbreak ACGGAGCCGA\allowbreak GAGGCACGCC\allowbreak CCGAGAGGGG\allowbreak CGGCGCGAAA\allowbreak GCGACGCCCC\allowbreak GAAAGGGGCG\allowbreak AGCGGAGACG\allowbreak CACGCGGAAG\allowbreak CGAGGCGGUG\allowbreak AGAACCGCCA\allowbreak GCGGAGACGC\allowbreak AGGGCCGGGG\allowbreak ACGGCCCAGG\allowbreak GGAAACCCAG\allowbreak GCGAGAGCCA\allowbreak GGCGGGGAAA\allowbreak CCCGCCAGCG\allowbreak GAGACGCAGC\allowbreak GCCCGAGAGG\allowbreak GCGCAGCGGA\allowbreak GACGCACCGG\allowbreak AGACGGGGGC\allowbreak GCCGAAGGGC\allowbreak GCCAGCCGAG\allowbreak AGGCACGGGC\allowbreak GGAGACGCCC\allowbreak GACCGGAGAC\allowbreak GGAGGUGAGA\allowbreak ACCAGGGGCG\allowbreak AAGACGCCCC\allowbreak ACGGCGC}\par\smallskip
\textbf{52}\quad 0.74~s, 80~nt\\
\texttt{AGCACAAAGC\allowbreak CAAAGAAGGA\allowbreak AAAAAGCAAA\allowbreak AACGGAAAAG\allowbreak ACCAAAAAAC\allowbreak GAAAAAACGA\allowbreak AAAAAGCAAA\allowbreak AAAGCAAGAA}\par\smallskip
\textbf{53}\quad 17.5~s, 400~nt\\
\texttt{AGAAGAAGCA\allowbreak AAAAAAGACA\allowbreak GAAGAAAAAG\allowbreak CAAAAAAAAA\allowbreak CCAAAAAGAA\allowbreak AGGGAAAAAA\allowbreak AACCAGAAAA\allowbreak AAGCGUAAAA\allowbreak AGAAGGGAGA\allowbreak AAAAAGCAAA\allowbreak AAAAAACGGA\allowbreak AAAAAGGGCG\allowbreak AGAAAAAAGG\allowbreak GAAAAAAAGG\allowbreak CGGAAAAAGG\allowbreak GGAGAAAAAA\allowbreak GCGAAAAAAA\allowbreak AAGCAAAAAA\allowbreak AAACCGAGGA\allowbreak AAAACCGAAA\allowbreak AAAAAGGGAA\allowbreak GGAAAAGGGA\allowbreak AAAGAAGGCA\allowbreak AAAAAAAGCG\allowbreak AGAGGAAAGC\allowbreak CGAAAAAAAA\allowbreak GCAAAAAAAA\allowbreak ACCGAAAAAA\allowbreak GAGCAAAAAA\allowbreak AAACGAAAAA\allowbreak AAAGGCAAAA\allowbreak AAAAACCGAA\allowbreak GAGAAUCGAA\allowbreak AAAAAAGGGG\allowbreak AAAAAAAACC\allowbreak AAAAAAAAGG\allowbreak GGAAAAAAAA\allowbreak GCAAAAAAAA\allowbreak AUGGAGAAAA\allowbreak AAGCAAAAAA}\par\smallskip
\textbf{54}\quad 0.88~s, 92~nt\\
\texttt{CCCCGCCGCU\allowbreak AGAGCGGCGC\allowbreak GCCGAGAGGC\allowbreak GCGGGCCAAG\allowbreak AGGCCCCCCC\allowbreak CGAGAGGGGG\allowbreak CGCGCGGGAG\allowbreak CGCGGGCGCG\allowbreak GGAGCGCCGG\allowbreak GG}\par\smallskip
\textbf{55}\quad 0.98~s, 102~nt\\
\texttt{AAAAAGCCGC\allowbreak GGACCGAAAG\allowbreak GUGGGAAAAG\allowbreak GGGACCCGGA\allowbreak AAGGGAGCCC\allowbreak CCGAAAAAGG\allowbreak GAAAGGCCCC\allowbreak CCCCCCGACG\allowbreak GCAAAAAAGA\allowbreak AAAGAAGAAA\allowbreak AA}\par\smallskip
\textbf{56}\quad 1.36~s, 387~nt\\
\texttt{GGCCCGCGCG\allowbreak CACGCCCAGU\allowbreak CGGCAAAGGC\allowbreak GCGGGAGCCG\allowbreak CAAUCCGCCA\allowbreak GCCGGGCACG\allowbreak CCGGGCCGGC\allowbreak CGGAGACGGC\allowbreak CGGCCCGGCG\allowbreak AAUGCCACCG\allowbreak CGAGGCCGGG\allowbreak GCGAGAGCCC\allowbreak CGGCCAACCC\allowbreak GGGAUACCGG\allowbreak GAGCCCGGGG\allowbreak GCGGCGCGGC\allowbreak ACCGCGCAGA\allowbreak GCCCCGGCGA\allowbreak AAAGCCAGGG\allowbreak GCAAGGGCCC\allowbreak CGCCCGUCCG\allowbreak CCCGCCGCGA\allowbreak AAGCGGCGGG\allowbreak CGGACGAGGC\allowbreak GGGGCCCCAG\allowbreak CGCCCAAAAG\allowbreak AGGGCGGCCC\allowbreak GCGCCCGAGA\allowbreak GGGCGCGGGC\allowbreak AAAUGCGGAA\allowbreak ACGCGAAGCC\allowbreak GAGAGGGCGA\allowbreak CCGCACGGAC\allowbreak GCGAAGGCGC\allowbreak CGGCGGGCGC\allowbreak CGCCCGGCUA\allowbreak CGGGCAGGCG\allowbreak AGCGCAGCGA\allowbreak GGCCAAA}\par\smallskip
\textbf{57}\quad 0.77~s, 36~nt\\
\texttt{GCAGGUGAGG\allowbreak AAACACGAAA\allowbreak GACGUCAGGA\allowbreak AACAGC}\par\smallskip
\textbf{58}\quad 0.73~s, 111~nt\\
\texttt{GGGCAGCCCA\allowbreak CGCGAAAGCG\allowbreak AUAGGAAACU\allowbreak GGGGCGAGAG\allowbreak CCAGGGGAAC\allowbreak CCCAGCCUAA\allowbreak AGGCAGCGGA\allowbreak GACGCAGGGC\allowbreak ACCCGCAAAA\allowbreak AAGCGGGGGC\allowbreak GAGAGCCGCC\allowbreak C}\par\smallskip
\textbf{59}\quad 1.21~s, 68~nt\\
\texttt{AAGACCGUGG\allowbreak GACCGGCCGG\allowbreak GAACCCGGCC\allowbreak GCGCCGCGAC\allowbreak GCUGCGAUAG\allowbreak CAGCGUCGCC\allowbreak ACGAAAAA}\par\smallskip
\textbf{60}\quad 3.55~s, 105~nt\\
\texttt{GAAGGCGCGG\allowbreak CGAGCAGGUU\allowbreak GUUCGGUUGG\allowbreak UCGGGCGCAC\allowbreak CCGAGAGGGA\allowbreak GACCCCGCGC\allowbreak ACCGUCACGG\allowbreak GACCAGAAAG\allowbreak GCGCGAAGAA\allowbreak AAAAAAAAAA\allowbreak AAAAA}\par\smallskip
\textbf{61}\quad 0.95~s, 67~nt\\
\texttt{GAAAAGGACC\allowbreak AGCAAAGCAC\allowbreak CAAAGGAAAG\allowbreak GCCACCACCA\allowbreak GCAAAGCACC\allowbreak AAAGGAAAGG\allowbreak GGAAAAA}\par\smallskip
\textbf{62}\quad 1.26~s, 53~nt\\
\texttt{AGGGCCACCA\allowbreak GCAGCCAGAC\allowbreak ACGAAACGGG\allowbreak GUAGGAGCGA\allowbreak GGAGGGAAAA\allowbreak CCG}\par\smallskip
\textbf{63}\quad 1.19~s, 400~nt\\
\texttt{AAAAGCCCGG\allowbreak CGGAAGACCG\allowbreak AAAGGUCAAG\allowbreak GGGUAGACCC\allowbreak CAAGCGGGAG\allowbreak ACCGCAAGCG\allowbreak CGAGAGCGCA\allowbreak ACCGCCGGGA\allowbreak AAAAGGGCCG\allowbreak CCAACCGCGG\allowbreak GAGCGGAACC\allowbreak GCGAGAGCGG\allowbreak AAGGGGGAAA\allowbreak CCCCACGGCG\allowbreak GGGACGCCAA\allowbreak GGCGGCCCGA\allowbreak AAAGCCCGCC\allowbreak CCACGCCGAA\allowbreak AGGCGAACCG\allowbreak CGAGAGCGGA\allowbreak AGGGCCAAAG\allowbreak CCCAACCGGG\allowbreak GAACCGGAAG\allowbreak GGCGGGCAAA\allowbreak AAGGGGGGGC\allowbreak AACGGCGAAU\allowbreak GCCGAAGCCG\allowbreak AAGACGGCCA\allowbreak GCGGGAGACC\allowbreak GCAACGCGGC\allowbreak AACGCGAAGC\allowbreak CCCCCCAAAA\allowbreak ACGGCGGGCG\allowbreak AGGGGGAAAC\allowbreak CCCAACGCCG\allowbreak AAAGGCGAAG\allowbreak UCCGAGAGGG\allowbreak CAACGCCGAG\allowbreak AGGCGAAGCC\allowbreak CGCCGAAAAA}\par\smallskip
\textbf{64}\quad 4.21~s, 110~nt\\
\texttt{GGAAAGCGAA\allowbreak ACCAAAAACC\allowbreak AAAAAAAAAG\allowbreak CGAAAAAAAA\allowbreak AAAAAAAAAA\allowbreak AAAAGAAAAA\allowbreak AGCGAAAAAA\allowbreak AAAAAAAGGG\allowbreak GAAAAAAAGG\allowbreak GGAAAAAAGC\allowbreak AAAGCCAAAA}\par\smallskip
\textbf{65}\quad 1.09~s, 40~nt\\
\texttt{AGGGUGGAAA\allowbreak AAGACGGUCA\allowbreak GAAGACCAGG\allowbreak CCCAAAAGAA}\par\smallskip
\textbf{66}\quad 0.76~s, 36~nt\\
\texttt{ACGGCCACAG\allowbreak GGCGCAAAGG\allowbreak GCAUGGGGGA\allowbreak AAGCGA}\par\smallskip
\textbf{67}\quad 0.95~s, 61~nt\\
\texttt{AAAAACGAAA\allowbreak AAAGGUAGCG\allowbreak GAAAACGCUU\allowbreak AAAAAAAGCA\allowbreak GAAAAAAAAA\allowbreak AAAAAAAAAA\allowbreak A}\par\smallskip
\textbf{68}\quad 2.63~s, 282~nt\\
\texttt{CCGGGAGACA\allowbreak CGGAGCCCGA\allowbreak AAGUGGCAGG\allowbreak GCGAGAGACC\allowbreak CACGGCGACA\allowbreak GACCGACGCC\allowbreak GAAAGAGCGA\allowbreak CGCCGAGAGA\allowbreak GCGACGCGGG\allowbreak UUCCGCCGCC\allowbreak GCCGCGGAAC\allowbreak GCGAGCGGCG\allowbreak GAGCCGCAAG\allowbreak CGGGGCCGGG\allowbreak ACCGGCGCAA\allowbreak AGCGCCGGGC\allowbreak CGGCCCCGCG\allowbreak GCCAGGGCCC\allowbreak AAAGGGCCCG\allowbreak GCCGCGGCAA\allowbreak ACGGGGCCCG\allowbreak CGAGCCCGAA\allowbreak AGAGGCAGGC\allowbreak CGAGAGAGCC\allowbreak AGGCGGAAAC\allowbreak AGCCAGGGGG\allowbreak AAACACCCAC\allowbreak GGCGAAAGAC\allowbreak CG}\par\smallskip
\textbf{69}\quad 6.89~s, 200~nt\\
\texttt{AGCGGGAAGA\allowbreak AGGCGGUAGC\allowbreak GAGGGGAAAA\allowbreak AAAGAAAAGA\allowbreak GAAAAAAAAA\allowbreak CCCACGGAUC\allowbreak CGACCAAACC\allowbreak GGCACCGGGC\allowbreak CCAAGGGAAA\allowbreak AAGGCGAAAG\allowbreak CAAAAAAACC\allowbreak GGGAAAAAAC\allowbreak GGAGGCGCGA\allowbreak AAAAGGGGGG\allowbreak GGUUCCGGUC\allowbreak CACCGGAAAA\allowbreak AAAAAAAAAA\allowbreak CGGAAGGGUC\allowbreak GGUUCCACCC\allowbreak AAGCAAGCCA}\par\smallskip
\textbf{70}\quad 10.5~s, 184~nt\\
\texttt{AAAGCGCCCG\allowbreak AAAACCGGCG\allowbreak GGGGCGAGCG\allowbreak ACGAAGCGAG\allowbreak GUGCGAGGCA\allowbreak GAGCACGUGC\allowbreak CAGCGCGGAG\allowbreak CGGCCACGAG\allowbreak GCGGGCGCAA\allowbreak AAAAAACGCC\allowbreak GCCAAAAUCA\allowbreak GCGAGGGCGG\allowbreak CGUGCCAGGG\allowbreak GACCUGCGGG\allowbreak GCAAAGCACG\allowbreak UGGGGCCGGG\allowbreak UGCGGGCCAC\allowbreak GGGAGGCGGC\allowbreak GGGA}\par\smallskip
\textbf{71}\quad 1.44~s, 88~nt\\
\texttt{GAAAGCGAAG\allowbreak GGAAAGGAGA\allowbreak AGCCAACGGU\allowbreak GCAAAGCGUC\allowbreak GGAAGGGAAA\allowbreak AAAUCAAAAA\allowbreak AGCCAUAAGA\allowbreak AAAGCAGAAA\allowbreak AAGUAAAA}\par\smallskip
\textbf{72}\quad 1.81~s, 73~nt\\
\texttt{AACGCCCGCU\allowbreak UAAAAAGCGG\allowbreak GGCGCGACAG\allowbreak CGCCCACGCC\allowbreak CGCGAGAGCG\allowbreak GGCGAGAAAC\allowbreak AAUAGCGGGC\allowbreak GAA}\par\smallskip
\textbf{73}\quad 842.9~s, 370~nt\\
\texttt{GGGCAAAGCC\allowbreak CGCCACCCGA\allowbreak AAGGCCCCCA\allowbreak AGCGGGUCGA\allowbreak AAGAGGGGCG\allowbreak CGCAAAGCGC\allowbreak AAGGCCAAAG\allowbreak GCCGCCCCGG\allowbreak GAAACCCCCG\allowbreak CAAGCCAGCC\allowbreak GUGAGGGGGG\allowbreak CGCGCAAAGC\allowbreak GCAAGGGCAA\allowbreak AGCCCGCCCC\allowbreak GGGAGACCCU\allowbreak GGCAAGCCCC\allowbreak GGGAAACCCC\allowbreak CGCGCCCAAA\allowbreak GGGCAAGGCC\allowbreak AAAGGCCGCG\allowbreak GGCCGAAAGG\allowbreak GGGGCAAGGA\allowbreak UGCGGAAACG\allowbreak GCGCGGGUCA\allowbreak AAGACCAACG\allowbreak GCAAAGCCGC\allowbreak GCGCGCGAAA\allowbreak GCCAUCCAAG\allowbreak GGACCCGAAA\allowbreak GGCCGGCCCC\allowbreak GAAACGGGAA\allowbreak GGGCAAAGCC\allowbreak CGCCGGGGGA\allowbreak GACCGUCCCA\allowbreak AGGGGGGGGA\allowbreak AACCGUGGCC\allowbreak CGCAAAGCGG}\par\smallskip
\textbf{74}\quad 296.2~s, 380~nt\\
\texttt{GUGCUAAAAG\allowbreak GGUCCAAAAG\allowbreak CUUCGAAAAG\allowbreak UUCCUGCGAA\allowbreak AACGUGCCCG\allowbreak GAAAAAGGGU\allowbreak CGGGGGUCCA\allowbreak AAAGGGGGCG\allowbreak UCGUUCCAAA\allowbreak AGCAUGCUGG\allowbreak GAGGGGUCCA\allowbreak AAAGGGUGGG\allowbreak ACGGUCGGGG\allowbreak GAAAACGUUG\allowbreak GUGGGUUCUU\allowbreak UCGUUCGAAA\allowbreak AGCUGCCGGG\allowbreak CAUAGUGGCG\allowbreak UCGAAAAGGU\allowbreak GCGUGCAGAG\allowbreak GUUCGGUCCG\allowbreak AAAAAGGUGC\allowbreak GUUCCGGUGU\allowbreak GCGGUCCCGG\allowbreak UGCUGCUUAG\allowbreak CGAGGGUGCC\allowbreak GUUCGGGAGU\allowbreak UCCUGCCUUC\allowbreak GCCAUCGGUC\allowbreak GGUCCGUCCG\allowbreak GGUCCAACCU\allowbreak GGCUGGCGGU\allowbreak UCGGUGCGGC\allowbreak CGGGUCCAGG\allowbreak GUUCUCUAGG\allowbreak UGCGCGUGGG\allowbreak UUCCGUUGCG\allowbreak GGUCCAGUGC}\par\smallskip
\textbf{75}\quad 0.85~s, 75~nt\\
\texttt{AAAAGACCGA\allowbreak AAAGGGCCCG\allowbreak GGAAAAGCCG\allowbreak AAAAAAAGGC\allowbreak CAAGCAAAAA\allowbreak GGGCGAAAAA\allowbreak ACGGAAAAAA\allowbreak AAGAG}\par\smallskip
\textbf{76}\quad 1{,}123.8~s, 393~nt\\
\texttt{GGGGCGAAAG\allowbreak CAGGCGCCGA\allowbreak AAGGCAACCG\allowbreak GACCGCAAGC\allowbreak CGAGAGGCGC\allowbreak CACCGAAAGG\allowbreak CCCAAAGGGG\allowbreak GGAAACCAGG\allowbreak CGCGGAAACG\allowbreak CAAGCGGACG\allowbreak AGAAGGGGAG\allowbreak ACCCGCCACC\allowbreak GAAAGGCCCA\allowbreak AAGGGGGGAA\allowbreak ACCAGGCCGC\allowbreak GAAAGCGAAC\allowbreak UCGAGGGCAA\allowbreak GGCGAAAGCC\allowbreak CCCAACGAAA\allowbreak GUCCCAAAGG\allowbreak GGGGAAACCA\allowbreak GGGCCGGAAA\allowbreak CGGAAGCCCA\allowbreak GCGGAACCCG\allowbreak AAAGGGGGCA\allowbreak CGGAGACGCC\allowbreak CAAAGGGGGG\allowbreak AAACCAGCCG\allowbreak CCGAAAGGCA\allowbreak ACCGCACCCC\allowbreak AAGCGGAGAC\allowbreak GCGGCACGGA\allowbreak GACGCCCAAA\allowbreak GGGGGGAGAC\allowbreak CAGCCGGGGA\allowbreak AACCCAAGGG\allowbreak GACCGGAAGC\allowbreak GGAAACGCGC\allowbreak CACCGAAAGG\allowbreak CCC}\par\smallskip
\textbf{77}\quad 1.12~s, 105~nt\\
\texttt{AAAAAGCCGG\allowbreak GGACCCCCGA\allowbreak AAGGGGCGCC\allowbreak CCGGGACGAA\allowbreak CGCCGACAGG\allowbreak CGAGGCCCCG\allowbreak AAAAAGGGGC\allowbreak AACGGCCCGG\allowbreak CCGGCAAAAA\allowbreak GAAGAAAAAA\allowbreak AAAAA}\par\smallskip
\textbf{78}\quad 512.0~s, 284~nt\\
\texttt{GUGGCCUGGC\allowbreak CGGAGCAAAA\allowbreak GUUGGGGCGA\allowbreak AAGCAAAACC\allowbreak UGGCGCGACG\allowbreak AAAAGCUGGG\allowbreak GGUGGGAACU\allowbreak UGCUGGUGAC\allowbreak AAAACGGUCC\allowbreak CGGGACAAAA\allowbreak GGGAAACCGU\allowbreak UUCGAAAAGG\allowbreak GUCGGUUGCC\allowbreak GUUGGGAACG\allowbreak GAAACGGGUG\allowbreak CCGGUGGGAA\allowbreak AACGUAGCCC\allowbreak GAAAGGAAAA\allowbreak GCCGCGCCUG\allowbreak CCAAAAGGUG\allowbreak CCCCUGACAA\allowbreak AGUGAGCGUG\allowbreak CGAAAAGGUG\allowbreak CCGCUGGGAA\allowbreak AACCGAAAGG\allowbreak CCUGGCAAAA\allowbreak GGUGCCCGUG\allowbreak CGCCUGCCAA\allowbreak CUGC}\par\smallskip
\textbf{79}\quad 1.97~s, 101~nt\\
\texttt{GAAAAACGCG\allowbreak GCGCGGCCAG\allowbreak AGGAACCCGA\allowbreak CGAAAGACGG\allowbreak GGAACGCCGA\allowbreak AGGAAAAAUC\allowbreak GCCGAAAGGC\allowbreak GCCAGCCGAG\allowbreak AAAAAAAAAA\allowbreak AAAAAAAGGA\allowbreak A}\par\smallskip
\textbf{80}\quad 33.8~s, 397~nt\\
\texttt{GAAAUAGGUC\allowbreak GAGCGCGCAA\allowbreak AGGCGCCAGG\allowbreak CGCGGAAACG\allowbreak GCCAAACGAG\allowbreak CGACGCCGAA\allowbreak GACCGGCGAA\allowbreak GGCCGAAGCC\allowbreak CGGGACCCCG\allowbreak AAGCCCCCCG\allowbreak ACCCGGAACG\allowbreak GGCGACGCGC\allowbreak CGGGGCCGGG\allowbreak GCGGGAGGGC\allowbreak GGAGCCCGGA\allowbreak CGGCGACGGC\allowbreak CCGUCGGAAA\allowbreak GCGACGAAAG\allowbreak AGGCCGGGAA\allowbreak ACGCCGAGAA\allowbreak ACCGGGGAGA\allowbreak GCCGCCGAAA\allowbreak GCCCGCGGAA\allowbreak ACGGCCGAAA\allowbreak GGGCGCCAAA\allowbreak GGCCCGGAAA\allowbreak ACCGGGGGAA\allowbreak UGGGGGAAAA\allowbreak GCGGGGAAAA\allowbreak ACCGGGGAAA\allowbreak ACGGCCGGGG\allowbreak GGCCGGAGAG\allowbreak GCGGCGCGAG\allowbreak AGCUCGGAGG\allowbreak GGGCCGGAAA\allowbreak GCCGCGGAAA\allowbreak UGCGCCACAA\allowbreak GGCGCGGAAA\allowbreak GGACCUUAAA\allowbreak GCAAAAA}\par\smallskip
\textbf{81}\quad 64.3~s, 212~nt\\
\texttt{AAAAGGCCAA\allowbreak CCCCCAACCG\allowbreak GAACCGCAAA\allowbreak AAGCGGCCGG\allowbreak GCAAAGCCAC\allowbreak GGCCGGGCCC\allowbreak CGAGCACGAG\allowbreak GGGCGAAAGC\allowbreak ACCCCGGCCC\allowbreak CCCAGGAGGA\allowbreak GCGAGAGCCC\allowbreak CCGCGGACCC\allowbreak GAAAGGCGCG\allowbreak CGCAAGGAGG\allowbreak GCGGAGGCGG\allowbreak GGGGGCAGCG\allowbreak AGCCACGGCG\allowbreak AGAGCCGGGC\allowbreak CGCGCCCCGC\allowbreak GAAACGCAGG\allowbreak AGGCAGCCGG\allowbreak CC}\par\smallskip
\textbf{82}\quad 1.12~s, 214~nt\\
\texttt{AAAGGGGGGU\allowbreak AGCCCCCCCG\allowbreak GGGCGAGAGC\allowbreak CCCCCCGGGG\allowbreak AAACCCGGGC\allowbreak CGGGGAGACC\allowbreak CGGGGCCGGG\allowbreak AGACCGGCCC\allowbreak CGGCGAGAGC\allowbreak CGGCGCCGGG\allowbreak AACGGCGGGG\allowbreak GCCGGGAGGC\allowbreak CCCACGUGCG\allowbreak AGAGCACGCG\allowbreak GGACGAGAGU\allowbreak CCCGCCGCCG\allowbreak AAAGGCGGCC\allowbreak GGGGGAGAUC\allowbreak CCGGGCGGGG\allowbreak ACACCCGCGG\allowbreak CCGCGAGAGC\allowbreak GGCC}\par\smallskip
\textbf{83}\quad 0.97~s, 119~nt\\
\texttt{AAAAGGGCCC\allowbreak CGCACCGAGA\allowbreak GGGCGAAAGC\allowbreak AGCGACCGAG\allowbreak GGAAACCGGG\allowbreak AGACCACGGG\allowbreak GGGAGGGAUC\allowbreak CCGCGAGAGC\allowbreak ACCCAGGGAG\allowbreak GGAAACCCCG\allowbreak AAAGGACCCA\allowbreak GGCCAAAAA}\par\smallskip
\textbf{84}\quad 2.86~s, 104~nt\\
\texttt{AAAAACCGGC\allowbreak CCGUCGAAAG\allowbreak ACCCGAAAAU\allowbreak GCGCGGGAAA\allowbreak GCGGAAAACG\allowbreak CCCCGGGGGA\allowbreak AAAACCCACG\allowbreak CACGGGGGAA\allowbreak CCGGAAAAAA\allowbreak AAAAGAAAAA\allowbreak AAAA}\par\smallskip
\textbf{85}\quad 508.8~s, 389~nt\\
\texttt{GCCGGGGGCC\allowbreak CGGGCAGCGG\allowbreak CAGCGAAAAG\allowbreak CAGCGAAAAG\allowbreak CACCGAAAAG\allowbreak GAGGGAAAAC\allowbreak CAGGGAAAAC\allowbreak CAGCCGCAGG\allowbreak CGAAAAGCCA\allowbreak CGCGCAGCGA\allowbreak AAAGCAGCGA\allowbreak AAAGCACCGA\allowbreak AAAGGAGGGA\allowbreak AAACCAGGGA\allowbreak AAACCAGCGC\allowbreak GACCGGAAAA\allowbreak CGGACCGUCA\allowbreak GCAAAAAGCA\allowbreak GCGAAAAGCA\allowbreak CCGAAAAGGA\allowbreak GGGAAAACCA\allowbreak GGGAAAACCA\allowbreak GACGGACCCG\allowbreak AAAAGGGAGC\allowbreak CCCAGCGAAA\allowbreak AGCAGCGAAA\allowbreak AGCAGGGAAA\allowbreak ACCAGGGAAA\allowbreak ACCACCGAAA\allowbreak AGGAGGGGCA\allowbreak GGGGAAAACC\allowbreak CAGGGCCAGC\allowbreak GAAAAGCAGC\allowbreak GAAAAGCACC\allowbreak GAAAAGGACC\allowbreak GAAAAGGAGG\allowbreak GAAAACCAGG\allowbreak CCCAGCCCGG\allowbreak GCCCCCGGC}\par\smallskip
\textbf{86}\quad 68.2~s, 355~nt\\
\texttt{AAAAAAACGC\allowbreak CACCCCAGGG\allowbreak ACGCGAAAGC\allowbreak GAGCCAGGGG\allowbreak AGACCCAGGC\allowbreak AGCGAGAACG\allowbreak CACCCAGCGG\allowbreak AAACGCAGGC\allowbreak GAAAAAAGCC\allowbreak AGGGAGCCAG\allowbreak GCACCCAGAA\allowbreak ACAGGGGAGG\allowbreak GCAAAGGAGG\allowbreak GAAGGCAAAG\allowbreak CAAAAAAAGC\allowbreak CCAGCGACGG\allowbreak GAGACCGGCG\allowbreak GAAACGCCCC\allowbreak GAGAGGGCGC\allowbreak ACCAAACGGC\allowbreak AGAAACAGCC\allowbreak AGCGACGGAC\allowbreak CGAGCGAAAG\allowbreak GCGGAGACGC\allowbreak ACGCACGCGA\allowbreak AAGCGACGGA\allowbreak GGCGAAAGCC\allowbreak ACCGAGCGGA\allowbreak GACGCACGCA\allowbreak GGGGGCGAGA\allowbreak GCCCGCGAAA\allowbreak GCGACCGGAG\allowbreak ACGGCCCAGG\allowbreak CAGCCGAAAG\allowbreak CCCAGGCGAA\allowbreak AAAGA}\par\smallskip
\textbf{87}\quad 13.2~s, 97~nt\\
\texttt{AAAAGCGCCG\allowbreak CGAAAAGCAC\allowbreak GGAAAAAAAA\allowbreak AAAAAAAAGA\allowbreak CACGCAAAGG\allowbreak GAGACCAAAC\allowbreak GGAGACGAAA\allowbreak AAAACGCAAA\allowbreak GAAGCCGCGG\allowbreak CGCAAAA}\par\smallskip
\textbf{88}\quad 0.96~s, 34~nt\\
\texttt{AAAAGCGCGG\allowbreak GCGCGAAAGG\allowbreak ACUCACGCGC\allowbreak AAAG}\par\smallskip
\textbf{89}\quad 21.9~s, 97~nt\\
\texttt{AAAAAAGCAC\allowbreak GGACUGCCAA\allowbreak CGAAGAGCGA\allowbreak AAGCCUGCCA\allowbreak ACGAAGCGGG\allowbreak GAGACCUACU\allowbreak GCCAAAGAAG\allowbreak GACGGCAAAA\allowbreak AAAAAAAAAA\allowbreak AAAAAAA}\par\smallskip
\textbf{90}\quad 1{,}397.5~s, 400~nt\\
\texttt{GGAAACACGA\allowbreak AAACAAGGGU\allowbreak CGGGUCGAAG\allowbreak CGAGGCAAAG\allowbreak CCCGAAAGGG\allowbreak CAGGCCGGAC\allowbreak GCACGGCGGA\allowbreak GCGAGAGCGC\allowbreak GACGGGGCGG\allowbreak GCGCGCCGGC\allowbreak GCCUGCGGCC\allowbreak GCCCGGACGG\allowbreak CAAAGCCGGC\allowbreak GGCGGGCCCA\allowbreak AGGCCGCGGC\allowbreak GGUCCCGCCC\allowbreak GAGGGCGCGG\allowbreak CGGCUCGAGA\allowbreak GUGGAGGCGG\allowbreak CCAGCCAAAG\allowbreak CCACCCUCAC\allowbreak GCGCCACGCG\allowbreak CCCACGCGCC\allowbreak CAGGACGCCG\allowbreak CGCGGGCCGC\allowbreak GCCGGCGCCC\allowbreak GGACCGGGGG\allowbreak CCGGAAACGG\allowbreak CCCGACGGCG\allowbreak CGGGACCCGG\allowbreak ACCGGCCGCC\allowbreak GCGGCGACGC\allowbreak GGGCAGGAGG\allowbreak GCAGGAGGCA\allowbreak GGAGAGGGAG\allowbreak CCGGAAACGG\allowbreak CAAAGGCACG\allowbreak CAAAGGUCGG\allowbreak GUCAAGGGAA\allowbreak GCAACGAAAG}\par\smallskip
\textbf{91}\quad 134.4~s, 392~nt\\
\texttt{CCAGGCCCAA\allowbreak AAAAAACGGC\allowbreak GUGGGUCAAA\allowbreak AAGGAGACAG\allowbreak GGGCCCCAAA\allowbreak AAAAAACCAA\allowbreak AAAGCCCCGA\allowbreak AAAGGGGAAA\allowbreak AAAGGAAACC\allowbreak UGCGAAAAAA\allowbreak AGUGGGAAAA\allowbreak ACUGCAAGGG\allowbreak ACCAAAAAGG\allowbreak AAACAGGGCA\allowbreak AAAAGCCAAA\allowbreak AAGUAGAGCA\allowbreak AAGGUCAGUG\allowbreak AAAAGGCAGG\allowbreak GGAAACGAAA\allowbreak AAACUGGGCC\allowbreak AAAGGACUGG\allowbreak AAAAAAAGGA\allowbreak GCCCCAAAAA\allowbreak AAAGCCCAAA\allowbreak AAGGCCGGAA\allowbreak ACCGGCCAGG\allowbreak GACCAAAAAA\allowbreak AAAAACGUGG\allowbreak GAAAAAAAAA\allowbreak GUGGGGAAAC\allowbreak GGGGCGGAAA\allowbreak ACAAAAAAAA\allowbreak GGAAAACGCC\allowbreak CCAACUGCAA\allowbreak AGGGGCCCCA\allowbreak AAAAAAAGUC\allowbreak CUGCGGCAAA\allowbreak AAGGAAAAAA\allowbreak CAUGGGCGGG\allowbreak GA}\par\smallskip
\textbf{92}\quad 5.80~s, 100~nt\\
\texttt{AAAAAAAGGA\allowbreak GGAGGACACC\allowbreak AGCAAAGCAA\allowbreak ACCCCAGCAC\allowbreak CAGGACACCA\allowbreak GCAAAGCAAA\allowbreak GGGCAGGACC\allowbreak AGGACACCAG\allowbreak CAAAGCAAAG\allowbreak GCCAAAAAAA}\par\smallskip
\textbf{93}\quad 0.93~s, 116~nt\\
\texttt{AAAAACUACG\allowbreak CCCCGCCCCG\allowbreak AGAGGGGAAC\allowbreak GCGGAGACGC\allowbreak GAACGGCUGA\allowbreak AGCCGCGGGG\allowbreak CGGAGGCGGG\allowbreak CAAAGCCCAA\allowbreak CGGGAAACCC\allowbreak GAAGCUCAAG\allowbreak GAGCGCCAAG\allowbreak AAAGGA}\par\smallskip
\textbf{94}\quad 440.3~s, 389~nt\\
\texttt{GGCAACCCGG\allowbreak CCGAAAGCCG\allowbreak CGCAAAGCGC\allowbreak GGAGGCCAAG\allowbreak CCGCGAACGC\allowbreak AAAGCGCGCA\allowbreak GGCCCAAACC\allowbreak GGGAAACCGG\allowbreak GAAGGGGCCA\allowbreak AGGCAAGGGA\allowbreak AGGGGGGGAA\allowbreak AAACGGGCGA\allowbreak AAGCCCGGAA\allowbreak AAGCCCGGCC\allowbreak CCCGCGAAAC\allowbreak GCGGCGCGGA\allowbreak AACCGCGGGG\allowbreak CGAGCGAUUG\allowbreak GAAUUUCGAG\allowbreak GUGCCGGAAA\allowbreak CGGCGUCAGG\allowbreak GACUGAGGCC\allowbreak CCGAAAGGGA\allowbreak AUCGCCCAAA\allowbreak GGGCCCCCCA\allowbreak GGCGCACGGC\allowbreak AAAAAAGCCC\allowbreak GCGAAAAAAA\allowbreak ACCGCAAAAA\allowbreak GCGGCGCGAA\allowbreak AGGCGGCCGC\allowbreak CCAGAAAAAG\allowbreak GGCGCGGGAA\allowbreak ACCGCGCAAC\allowbreak CCGCAAAGCA\allowbreak GCGGCAAAGC\allowbreak CGCAAGGGCC\allowbreak GAAAUGGGGC\allowbreak GCCAAAGCC}\par\smallskip
\textbf{95}\quad 1{,}167.7~s, 82~nt\\
\texttt{AAGCGGAGCC\allowbreak CCCCGGAAAC\allowbreak GGGGACCGCG\allowbreak AAAGCGGUGG\allowbreak CCCGGCCCGG\allowbreak AAACGGGUGC\allowbreak CCCGAAAGGG\allowbreak GUCGGUCCGC\allowbreak AA}\par\smallskip
\textbf{96}\quad 359.0~s, 358~nt\\
\texttt{CUGGAAGGAA\allowbreak ACAAGCGUGC\allowbreak AGGGGCAAAG\allowbreak CCGACGAGUG\allowbreak CGCAAGUGCG\allowbreak AAAGUGCACU\allowbreak GGGAGACUGG\allowbreak AAAAAGGAAA\allowbreak CAACUGGAAG\allowbreak UGGAAGGAAA\allowbreak CAAGGCAACA\allowbreak GGGCGAAACG\allowbreak CGACGAGGAG\allowbreak CCAACUGCGA\allowbreak AAGUGGAGUG\allowbreak CGAAAGUGCA\allowbreak AAAAGGAAAC\allowbreak AACUGCAACC\allowbreak UGCAAGGAAA\allowbreak CAACCCUGGA\allowbreak GGGCGAAACG\allowbreak CGACGACUGG\allowbreak GGAAGGUGGA\allowbreak AACGUCACUG\allowbreak GGAGACUGGA\allowbreak AAAAGGAAAC\allowbreak AAGUGGGAAG\allowbreak GUGGAACGAA\allowbreak AGAAGCCUGG\allowbreak AGACCGAAAC\allowbreak GGAGCGACUG\allowbreak GGCAAGGUGG\allowbreak AAACGUCACU\allowbreak GCGAAAGUGG\allowbreak AAAAACGAAA\allowbreak GAACUGCC}\par\smallskip
\textbf{97}\quad 1{,}188.8~s, 400~nt\\
\texttt{ACGUCCAAGC\allowbreak CCGAAAACCG\allowbreak CGAAAAAACG\allowbreak CGGCGGGCCC\allowbreak GGAAAGGCCG\allowbreak AAAAGGCCCC\allowbreak GGGGGAAGGG\allowbreak GAGACCCCCC\allowbreak AGCGAAACCC\allowbreak GAAAGGCCGA\allowbreak AAAAGGCCCG\allowbreak GGGGCAAGCG\allowbreak GAAAACGCGC\allowbreak CGGAGCGAAA\allowbreak GCCCAGGGGA\allowbreak AAAGCGAACC\allowbreak GGAAAAACGG\allowbreak CGCGGACCGA\allowbreak AAAGGCCGGC\allowbreak GAGAGCCAGC\allowbreak CGGAAAAACG\allowbreak ACCAAAAAAG\allowbreak GCGGCCGAAA\allowbreak AGGCGCGAAA\allowbreak GCACCCCGCA\allowbreak AAAAACGCGA\allowbreak AAAAGCGGCG\allowbreak AAAAGCAAAC\allowbreak CCCCCGAGAC\allowbreak GGGGGGAAAA\allowbreak AAAGCGGGGA\allowbreak AAAAACCGGC\allowbreak AAAAACCCCA\allowbreak AAACGCAAAG\allowbreak GUUCAAAGGC\allowbreak GAGGCCAAAA\allowbreak GCGAGAGCGC\allowbreak GAAAGCGCCG\allowbreak CGAAACGCGG\allowbreak CAGGCCCGGC}\par\smallskip
\textbf{98}\quad 24.4~s, 202~nt\\
\texttt{AAAGGGGGCG\allowbreak AAACGCCAAG\allowbreak CCGAAACGGC\allowbreak AACGCCAAAG\allowbreak GCGCCCAAAA\allowbreak AAAAGCCGAC\allowbreak GAAACGUCAA\allowbreak GCCCAAAGGG\allowbreak CAAGGCCAAA\allowbreak GGCCGGCAAA\allowbreak AAAAACGCGG\allowbreak GCAAAGCCCA\allowbreak ACGCGAAACG\allowbreak CGAACCCGAA\allowbreak ACGGGGCGAA\allowbreak AAAAAACCCG\allowbreak CGCAAAGCGC\allowbreak AACCGCAAAG\allowbreak CGGAACCGGA\allowbreak AACCGGGGGA\allowbreak AA}\par\smallskip
\textbf{99}\quad 5{,}028.1~s$^{\dagger}$, 364~nt\\
\texttt{GCCGGCCCCC\allowbreak GGGCGGCGAA\allowbreak GACUGCCACG\allowbreak GCGCGCGCGA\allowbreak AAGACCGACC\allowbreak CGAGAGUGGC\allowbreak GAGCCGCCCA\allowbreak CCCGCUGGAC\allowbreak UGGCACGGAC\allowbreak CUGGACCUUG\allowbreak GAGCGCGAAA\allowbreak GACCUCGCAG\allowbreak GCGCAGCGCC\allowbreak GAGACUGGCU\allowbreak GGGGAAACAA\allowbreak CGGGGCCCGC\allowbreak CGCAAAGCGG\allowbreak CACGCCCAAA\allowbreak GGGCGGGGGA\allowbreak AACGCAGGGC\allowbreak AGGGGAGCGC\allowbreak AGGGCAAAGC\allowbreak CGGGCAAAGC\allowbreak CCGGCAGGGG\allowbreak CAAAGCCCCG\allowbreak CCCGCCCCGC\allowbreak AAAGCGGGGC\allowbreak AGCCGCAAAG\allowbreak CGGCGCGCGG\allowbreak GCAAGGAAAC\allowbreak CUGGGAGUGG\allowbreak ACGUAUUGGA\allowbreak GCGCGAAAGA\allowbreak CCUCGCAUAG\allowbreak GGACUGCACU\allowbreak GGGCCCAGCC\allowbreak GGCA}\par\smallskip
\textbf{100}\quad 389.2~s, 382~nt\\
\texttt{AAAAAACCGA\allowbreak AAAAAUCAGG\allowbreak GAAAAAAGCA\allowbreak GGGAAAAAAA\allowbreak AAAAAAACCA\allowbreak AAAAAAAACC\allowbreak GAAAAAAAAA\allowbreak AAAAAGGAGC\allowbreak GAAAAAACCA\allowbreak GGAAAAAAGC\allowbreak CGAAAAAAAA\allowbreak AAAAAAGGAA\allowbreak AAAAGCCACC\allowbreak GAAAAAGCCA\allowbreak GGGAAAAAAA\allowbreak AAAAAAACCA\allowbreak AAAAAAAACC\allowbreak GAAAAAAAAA\allowbreak AAAAAGGAGG\allowbreak AAAAAAAGGA\allowbreak CCGAAAAAAC\allowbreak CGAAAAAAAA\allowbreak AAAAAAGGGA\allowbreak AAAAAGGAGC\allowbreak GAAAAAACCA\allowbreak GCGAAAAAAA\allowbreak AAAAAAAGCA\allowbreak AAAAAAAAGG\allowbreak GAAAAAAAAA\allowbreak AAAAACCAGG\allowbreak GAAAAAAGCA\allowbreak GCGAAAAAAG\allowbreak GGAAAAAAAA\allowbreak AAGCCGAAAA\allowbreak AAAAAAAAAA\allowbreak GGACCGAAAA\allowbreak AAGCAGAGAA\allowbreak AAAAGGAAAA\allowbreak AA}\par\smallskip
\endgroup

\end{document}